\documentclass[prb]{revtex4}
 
\usepackage{graphicx}
\usepackage{dcolumn}
\usepackage{bm}
\usepackage{color}

\usepackage{booktabs}
\usepackage{longtable}
\usepackage{makecell}
   \newcolumntype{C}[1]{>{\centering\arraybackslash}p{#1}}
 \AtBeginDocument{
\heavyrulewidth=.08em
\lightrulewidth=.05em
\cmidrulewidth=.03em
\belowrulesep=.65ex
\belowbottomsep=0pt
\aboverulesep=.4ex
\abovetopsep=0pt
\cmidrulesep=\doublerulesep
\cmidrulekern=.5em
\defaultaddspace=.5em
}

\newcommand{\bx}{{\bm x}}

\graphicspath{{./figures/}} 
  
\begin{document} 

 
\title{Exploring Critical Points of Energy Landscapes: From  Low-Dimensional Examples to Phase Field Crystal PDEs}
 
\author{P.~Subramanian}
\affiliation{Mathematical Institute, University of Oxford, Oxford OX2 6GG, UK}

\author{I.G. Kevrekidis}
\affiliation{Department of Chemical and Biomolecular Engineering, Johns Hopkins University Whiting School of Engineering}

\affiliation{Department of Applied Mathematics and Statistics, Johns Hopkins University Whiting School of Engineering}

\author{P.\,G. Kevrekidis}
\affiliation{Department of Mathematics and Statistics, University of
  Massachusetts, Amherst MA 01003-4515, USA}
\affiliation{Mathematical Institute, University of Oxford, Oxford, OX2
  6GG, UK}
  

\begin{abstract} 
In the present work we explore the application of a few root-finding methods to a series of prototypical examples. The methods we consider include: (a) the so-called continuous-time Nesterov (CTN) flow method; (b) a variant thereof referred to as the squared-operator method (SOM); and (c)  the the joint action of each of the above two methods with the so-called deflation method. More ``traditional'' methods such as Newton's method (and its variant with deflation) are also brought to bear. Our toy examples start with a naive one degree-of-freedom (dof) system to provide the lay of the land. Subsequently, we turn to a 2-dof system that is motivated by the reduction of an infinite-dimensional, phase field crystal (PFC) model of soft matter crystallisation. Once the landscape of the 2-dof system has been elucidated, we turn to the full PDE model and illustrate how the insights of the low-dimensional examples lead to novel solutions at the PDE level that are of relevance and interest to the full framework of soft matter crystallization.
\end{abstract} 

 
\maketitle 

\section{Introduction and Motivation}

The topic of root finding at the level of stationary states of ordinary and
partial differential equations is one of fairly universal appeal~\cite{kelley,bertini}.
Especially in the realm of nonlinear waves and pattern formation, a wide variety
of methods has been proposed including in the recent past~\cite{3,2}, many
of which have been summarized in~\cite{yangbook}. These types of methods
enable the identification of steady patterns, which are subsequently subject
to the evaluation of their respective stability properties. When the associated
patterns are found to be stable,
they suggest natural attractors of the dynamics (in the case
of dissipative dynamical systems) or centers around which the dynamics
may oscillate (in conservative dynamical systems). 
On the other hand, even unstable dynamical solutions (potential maxima
or saddles of the landscape of the system) are interesting in their own
right towards characterizing transient  dynamics and transitions between
different states.

In the past few years, there has been a number of proposals for variations
to the well-known gradient flow methods in order to take advantage
of the acceleration that may be provided by the so-called Nesterov method~\cite{nesterov}. While the original method was proposed
for discrete systems, it has been recently adapted to continuum
ones both at the ODE level~\cite{candes} and at the PDE level~\cite{cory}.
It has also been shown to be suitable to frame in the realm of 
Lagrangian and action extremization principles~\cite{MJ} and has proved
to be an efficient method for identifying steady states of even complex
multi-component PDE systems (in connection to spinor atomic Bose-Einstein
condensates) in~\cite{christian}.

Our aim in the present work is to expand upon the relevant ideas,
using a few simple examples which will provide a useful understanding
about how these methods work, especially so in a progressively more
complex landscape starting with 1-dof, progressively
moving to 2-dof's and ultimately connecting the latter with a full PDE. That is to say, our aim is to explore
the relevance of the methods in well-controlled ODE and, subsequently,
in (connected to the ODE through suitable reductions) PDE
examples as a means of identifying some of the advantages and potential
caveats of different methods and as a way to guide the selection of
which tool is most well-suited to which example.

Finding equilibria of a system of coupled ODEs is simplified if the corresponding conditions for attaining equilibrium reduce to a set of coupled polynomials/Laurent polynomials. In that case, multiple iterative local methods such as Newton's method or non-iterative global methods can be employed. Local methods need a good initial guess and a method to either determine the Jacobian or its action on a vector. Alternatively, global methods such as homotopy use ideas from numerical algebraic geometry to obtain all non-singular solutions (both real and complex) to the set of coupled polynomial equations without any initial guess~\cite{bertini,sommesebook}. Finding equilibria of the PDE system using iterative methods is subject to the same constraints as described above in the discussion of ODE systems, in that there is a need for a ``good'' initial guess and the need to approximate the action of the Jacobian. One way to generate suitable initial guesses is to (i) use asymptotic states obtained from the time evolution of the full PDE model, or (ii) generate a reduced order ODE model that reproduces the dynamics of the PDE model in the neighbourhood of a critical point. Such a reduction is at the heart of a wide variety of techniques at the level of perturbation theory, effective Lagrangians/Hamiltonians, variational methods and many others; see, e.g., the reviews of~\cite{kivshar,variational,kivsharmal} for just a few examples. This second approach utilises appropriate rescaling of the variable(s), key parameters and time in the PDE model, to derive effective equations
chracterizing the transition at the critical point. Once we obtain a reduced set of ODEs, the different ODE root-finding methods described above can be deployed to obtain the associated equilibria for the ODE system and then to transcribe those towards initial guesses of the relevant PDE.

In the exposition below we start with a very simple 1-dof example where the solution landscape is fully-known, as the roots are transparent from the very form of the ODE's right hand side (RHS). This allows us to easily transcribe the CTN method in this context and observe its potential convergence to different types (single, and double) roots. This also simplifies the application of modified methods, such as the method that we will refer to as the
square-operator method (SOM)~\cite{yangbook,cory} which does effectively use the Jacobian, retaining the form of a flow method but in a new landscape where the original roots are now landscape minima (and hence each one
of them is locally attractive --or for conservative dynamics,  a center--). An additional widely used methodology in the form of the so-called deflation method~\cite{farrell2,farrell3,egc_16,egc_20} is also brought to bear. This method focuses on eliminating the possibility of the flow to converge to a particular root. As a byproduct, the use of root finding techniques (such as Newton's method), in conjunction with the repeatedly-modified landscape induced by sequentially ``removing'' from the pool of available attractors different roots, can offer
convergence to numerous distinct roots. We explore how deflation works both with flow methods and with Newton type methods in our simple examples.

Indeed, in all of these methods, we go beyond the informative but rather limited 1-dof example and explore an equally tractable, but more complex 2-dof example. This setting has emerged from a 2-dof reduction (near a suitable critical point as described above) of an infinite dimensional model of the PFC variety exploring the crystallization of soft matter~\cite{SAKR16,priya_ref}. For the case considered, we find a significant wealth of possible stationary states (9 in total), covering different
types of states: energy minima, energy maxima
and saddle points. While naturally root-finding techniques can identify all 9 of these roots, here we explore the behavior of different types of flow methods, such as CTN, SOM, deflation-based CTN, etc. We examine what
converges where, what happens to the landscape under deflation, what are the basins of attraction of each root, etc. All of these questions reveal the strengths and weaknesses of the different methods. Arguably, even more importantly, these set the stage for reconstructing, on the basis of the 2-dof ODE steady state, viable PDE solutions of the original system. We then explain the effort to capture the corresponding PDE steady states, the comparison
of the stability properties of the PDE vs. those of the ODEs and the potential of the instabilities of the states obtained (based on the reduction) to yield new asymmetric states (via the dynamical evolution), of not only quasi-crystalline but also other types.

Our presentation will be structured as follows: in section II, we will kick off by explaining the different methods in the simplest 1-dof example. Then, in section III, we will expand considerations to the example at the center of our analysis, namely the PFC model of crystallisation in soft matter. The first part of the relevant section will focus on the more definitive 2-dof considerations: upon performing the relevant reduction of the PDE
to the 2-dof model, we will explore the latter for specific parameter sets at some length. Then, we will return to the PDE to examine what solutions we can obtain from the evolution of the reconstructed (based on the 2-dof) PDE
solutions. A broad range of not only quasi-crystalline, but also rhombic, hexagonal and other asymmetric patterns will be seen to emerge from this effort that are of intrinsic interest in their own right for the original PFC system.
Finally, in section IV we will summarize our findings and offer some possibilities
for extensions of the present work.


\section{One-dimensional example}

We consider the dynamics of a system defined as
\begin{equation}
\dot{x} = f(x) = - (x-1) (x-3)^2 (x-6)= -\nabla V(x).
\label{eqn:1dsysrhs}
\end{equation}
Here the overdot indicates a derivative with respect to time and the corresponding energy landscape can then be defined as $V(x) = -\int f(x)\*dx$.
In order to determine the solution landscape in this simple example, the first method that we use is the continuous time Nesterov (CTN) method~\cite{candes}.
In this method, we consider the following ODE (i.e., the flow) in order to solve the problem $f(x)=0$,
\begin{eqnarray}
  \ddot{x} + \frac{3}{t} \dot{x}=f(x)\,.
  \label{eqn1}
\end{eqnarray}

One can detect two principal features that distinguish CTN dynamics from the standard
gradient descent flow: $\dot{x}=f(x)$. The dynamical system associated with CTN dynamics involves a second order and also 
non-autonomous ODE. As a result, the acceleration vector, rather than the velocity vector, 
is proportional to $f$ (at least for large $t$).  Considering a mechanical analogy, the relevant (fictitious) particle $x$ has been given a mass and a time-dependent dissipation, which has the unusual form: $({3}/{t})\,\dot{x}$. This effective damping on the fictitious particle is tuned to be large at the initial time, i.e., far from the equilibrium, while it becomes weaker as one (hopefully) approaches the relevant fixed point. This term is therefore responsible for the convergence towards one of the roots of $f$.

\begin{figure}[]
\centering
{\vspace{0.3cm}\includegraphics[width=12cm]{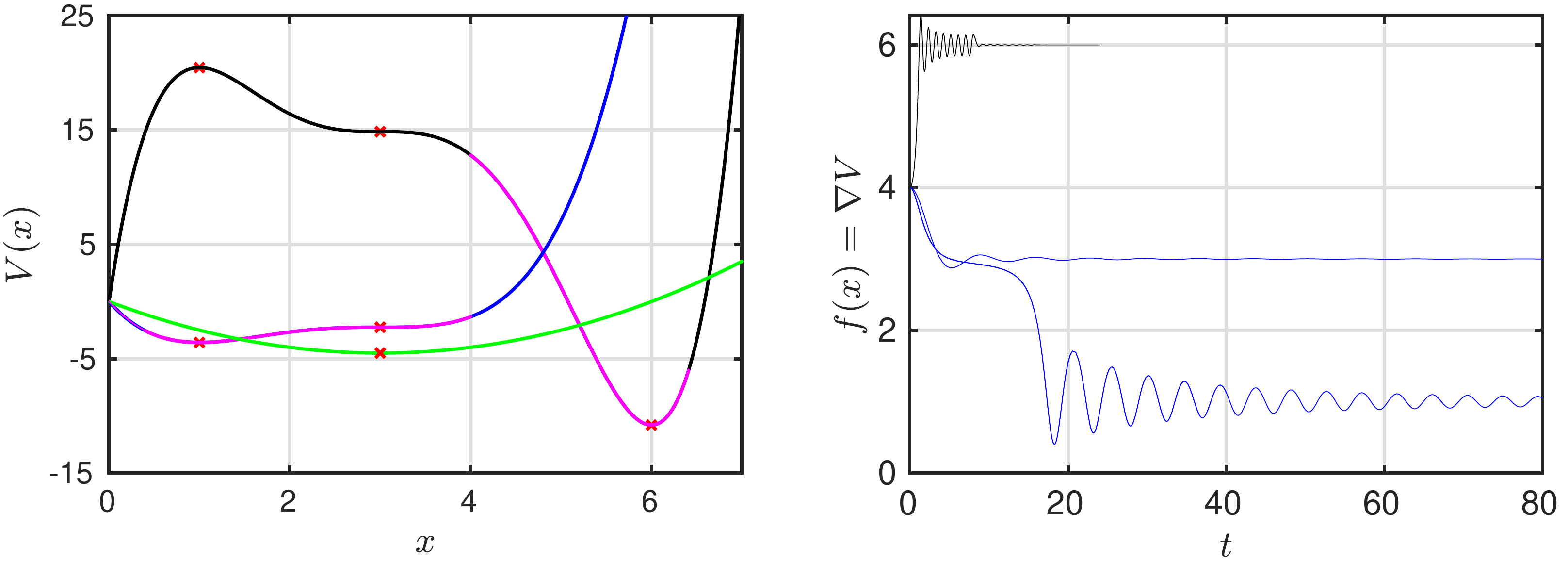}}
\caption[]{Left: Energy landscape for the one-dimensional example. Right: Continuous time Nesterov (CTN) iteration. Black lines in both panels is the original system with roots (1,3,3,6). Blue lines indicate the energy landscape and iterations after removing one root, i.e., the minimum at $x^*=6$. This leads to a new minimum being $x^*=1$. Green line is after removing $x^*=1,6$ and one of the repeated roots, i.e., $x^*=3$. }
\label{fig:1dexam}
\end{figure}
We now turn to the specific example of interest. The function $f(x)$ has three roots (one of which is double): a single root $x=1$, a double root at $x=3$ and another single root at $x=6$. Here the equilibrium at $x=6$ is the global minimum of the associated potential energy landscape $V(x)$, $x=1$ is the global maximum while the double root at $x=3$ corresponds to a semi-stable point as shown in Figure \ref{fig:1dexam} left panel. Here the energy landscape $V(x)$ is plotted as a function of $x$ in black. The equilibria are identified with red crosses. The right panel of the figure shows the  time evolution of the associated CTN flow. Starting from an initial condition, say, $x=4$, we see that the evolution (in black line in the right panel, pink overlay on the black line in the left panel) approaches the energy landscape minimum at $x^*=6$. In order to approach any other equilibrium, we need to modify the system. If we explicitly factor the known solution out of the system, i.e., define a new system $f_{new}=f/(x-6)$, we obtain the blue curve in the energy landscape. Notice that this is a rather {\it dramatic} change, more dramatic in a sense than the deflation considerations that will be given below, among other things because this changes the asymptotic behavior of the field at large $x$; it also alters the stability of $x=1$ which is now no longer a maximum, but
rather a minimum of the dynamics. This implies that an evolution starting from the same initial condition of $x=4$ will now approach the value of $x^*=1$ which is the new minimum. It is interesting to highlight the difference between CTN and a gradient descent here too. The latter, starting at $x=4$ would get attracted to the semi-stable
point $x^*=3$ without being able to reach $x^*=1$. Yet, the CTN enables an oscillatory motion (through its acceleration term) which, in turn, allows the system to bypass the semi-stable point and reach the, now stable,
$x^*=1$. 

We can carry this further and remove both roots $x^*=1$ and $x^*=6$ in the above rather crude way. Then we find that the CTN evolution is unable to asymptote to any solution (not shown here). Once again the oscillatory component of the method reveals the effective metastability of the root, rendering the method unable
to converge to it. When the double root nature of this point is taken into account (i.e., one factor of $x-3$ is removed
similar to before), we obtain the energy as shown in green  in the left panel. Starting from the same initial condition, CTN is finally capable of converging to the value of $x^*=3$, which of course has by now become a stable fixed point of the highly modified dynamics.

From the above, it becomes clear that CTN can only converge to the stable
equilibria of the flow in this one-dimensional example. In order to converge
to unstable equilibria, a drastic modification of the relevant landscape
(like the division by the monomial factor involving the already identified root,
here the $x-6$) is needed. Thus rendering $x=1$ a minimum,
it is also possible converge to the latter root. Semi-stable fixed points
appear not to be well-suited to CTN given its oscillatory nature and, thus, its exploration of
the unstable direction of the semi-stable equilibrium. A natural query that arises
is whether one can modify the method to be able to converge (for suitable
initial guesses) to all equilibria. To address this issue, we now
discuss the squared operator method (SOM).

A modification that allows us to convert all the equilibria of a given system $f(x)$ into minima is the SOM. Here we define a new system $g(x)$ such that 
\begin{equation}
g(x) = -f'(x)\* f(x)\,.
\end{equation}
All equilibria of the $f(x)$ system are now minima in this new system $g(x)$ (as $g'(x^*)<0$ at $f(x^*)=0$). Defining the corresponding $g(x)$ for the system defined in Eq.~(\ref{eqn:1dsysrhs}), we calculate the associated energy landscape. In this squared operator system we find that there are 5 equilibria at $x=1,1.7,3,5.15,6$, with the minima at $x=1,3,6$ and maxima at $x=1.7, 5.15$. Note that the maxima are {\it spurious} in connection to the original
root finding problem. They are ``topologically'' generated separatrices for the dynamics of the squared operator, created by the fact that we are now minimizing an effective potential energy $V_{eff}=\frac{1}{2} f(x)^2$. In this setting, the roots that we previously obtained (and only the roots of the original problem) are the global minima of the novel potential energy landscape $V_{eff}$. Hence, between these minima, it is topologically necessary that
(at nonvanishing values of $V_{eff}$ and thus of $f$), maxima of $V_{eff}$ will arise. These maxima now naturally also partition the space between the corresponding minima. In particular, initial conditions in the range $-\infty<x<1.7$ reach $x^*=1$, $1.7<x<5.15$ reach $x^*=3$ and $x>5.15$ reach $x^*=6$. These ranges are indeed the one-dimensional basins of attractions of each of the asymptotic minima in this system. 

However, in addition to obtaining different roots starting with initial data in different basin boundaries, an arguably more desirable scenario would be to obtain ideally all roots starting from a single initial guess in the
spirit of deflation methods. Thus at the next step, we attempt to combine CTN  and SOM with deflation methods. Below, we will show the case example of (the better behaved, between the two, as we saw above) SOM
to illustrate the ``pathologies'' present. We have identified similar pathologies in the CTN case.

\begin{figure}[]
\centering
{\vspace{0.3cm}\includegraphics[width=7.5cm]{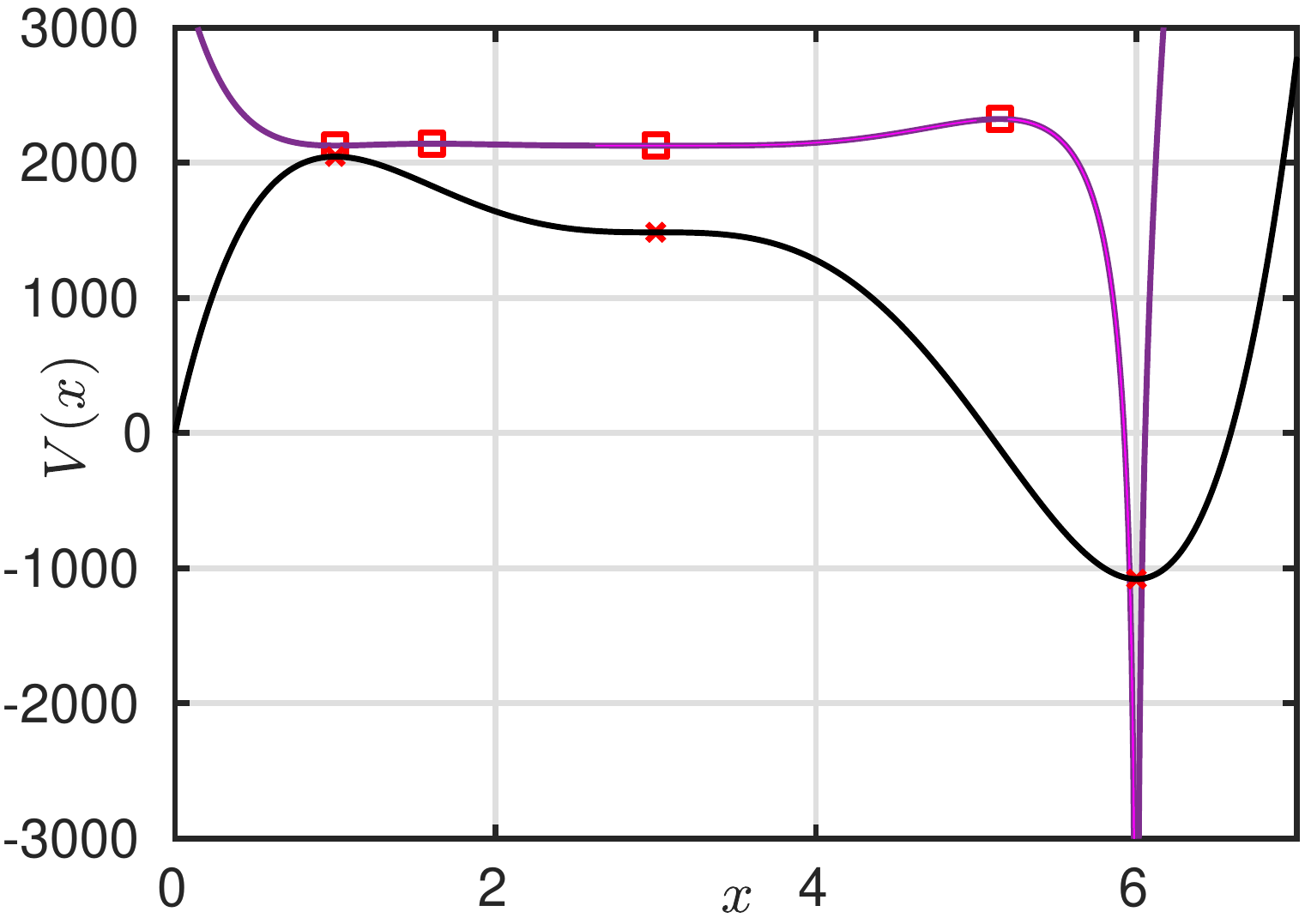}}
\caption[]{Energy landscape for the one-dimensional example and the related SOM with deflation. The purple curve is the energy associated with the squared-operator system with deflation, i.e., $G(x)$, where the potential energy obtained by the (negative) anti-derivative thereof. Red squares indicate the equilibria for the squared-operator dynamics. The black line is the energy associated with $f(x)$ (and its associated $V_{eff}$) shown here multiplied by a factor of $100$ (to bring it to the same scale as the other curve) where red crosses are the equilibria. }
\label{fig:1dexam2}
\end{figure}

In order to explore the relevant technique (SOM with deflation), having obtained an equilibrium of the system, say at $x^*=6$, we deflate this equilibrium to define a new system $G(x)$ as below,
\begin{equation}
G(x) = g(x)\times \left(\frac{1}{||x-x^*||^{p}}+\alpha \right).
\label{eqn:1ddefl}
\end{equation}
Generally we use a factor with squared norm, i.e., $p=2$ and a (positive)
shift operator $\alpha=1$ in the above factor. The corresponding energy landscape is shown in the Figure \ref{fig:1dexam2}
in purple, with the equilibria of this deflated squared operator system shown as red squares. We also show an exaggerated version of the un-deflated operator in black (multiplied by a factor of $100$) for comparison. Starting the time-marching from the initial value of $x=5.1$, the squared operator dynamics reaches the equilibrium at $x^*=3$. If however, we start the time-marching from the same point, e.g., $x=5.2$, the evolution still attempts to reach $x^*=6$, since the deflated squared operator in purple has continuously decreasing value for $5.15<x<6$. Normally the stopping criterion in a CTN method would be to require both the velocity $\dot{x}$ and acceleration $\ddot{x}$ to vanish. In this case, we observe that as the evolution approaches $x^*=6$, the slope of the potential is non-zero. To recognise that we have reached an equilibrium, a modified stopping criterion in this case would be to check if $g(x_{current})<1e-10$. This makes the evolution stop when we reach an equilibrium, even if the velocity $\dot{x}$, in the CTN formulation is non-zero. 

Such a modification of the stopping criteria is necessitated when combining deflation with a flow method, even well-behaved ones as in the SOM case utilized here. The reason is simple, a posteriori. Deflation eradicates one of the roots of the original landscape. In its stead, it leaves behind a pole (typically). This pole may form an infinitely deep potential well which, in turn, will render it impossible to get unstuck from the flow towards what used to be an equilibrium but is now merely a pole. It may be interesting to consider infinity-crossing methods in this regard (see for a recent example~\cite{yannis}). Nevertheless, the generic feature remains: the depth of the landscape around a former minimum may change, e.g., with its becoming deeper in the form of a pole. Yet, in the dynamics of the flow considered, the pole continues to attract our original initial condition and thus does not allow it to  to converge to another root as is the goal of the deflation technique.

\begin{figure}[]
\centering
{\vspace{0.3cm}\includegraphics[width=8cm]{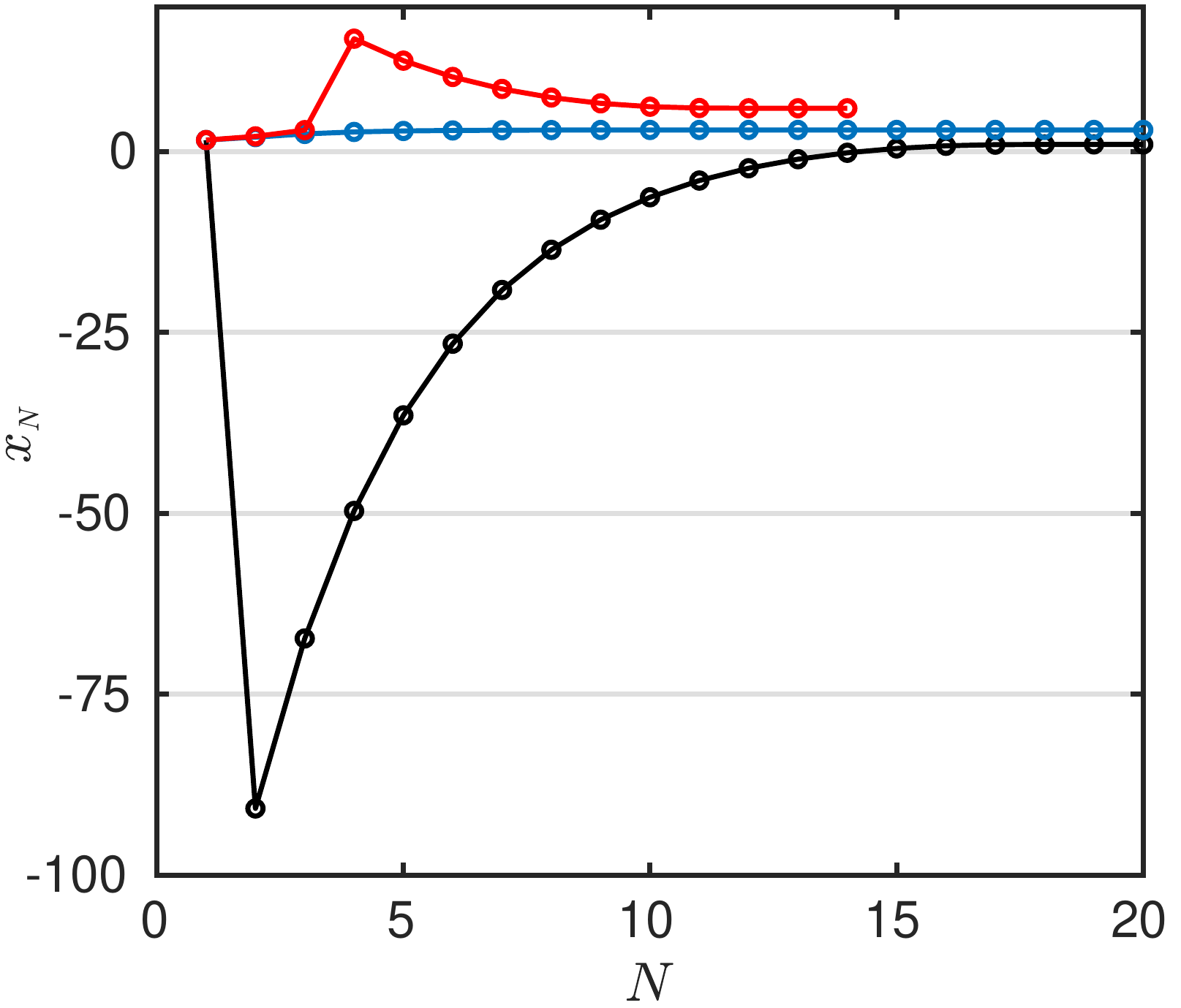}}
\caption[]{Newton with deflation for the system governed by $f(x)$ plotted as a function of the index of the Newton iteration $N$ and the current Newton iterate $x_N$. Starting from the same initial guess of $x=1.6$, we first converge to $x=1$ (black line with markers). We this solution deflated, we approach $x=3$ (blue line with markers). When both these solutions are deflated, we recover $x=6$ solution (red line with markers). }
\label{fig:1dexam3a}
\end{figure}

A final technique to employ is the well established~\cite{farrell2} combination of Newton iteration with deflation, for which we show the results in Figure \ref{fig:1dexam3a}. Here, the value of the Newton iterate $x_N$ is shown as a function of the number of iteration $N$. Starting from an initial choice of $x=1.6$ and using a Newton iteration for $f(x)$, we obtain the equilibrium at $x^*=1$ (black curve with circle markers).  When this equilibrium of the $f(x)$ dynamics is deflated in the same fashion as described in Eqn.\,(\ref{eqn:1ddefl}), the Newton iteration is able to converge to another equilibrium at $x^*=3$ (blue curve with circle markers). Deflating both of
these equilibria, allows the Newton iteration to approach the third
equilibrium at $x^*=6$ (red curve with circle markers). This confirms
that the combination of a Newton iteration with deflation yields far
more promising results.
For suitable initial guesses, the algorithm can escape  from the infinite negative (effective) potential energy pole formed at $x^*=1$ under deflation and ``jump'' outside of the relevant effective potential well through its discrete steps. Such a possibility enables it to retrieve additional equilibria and ultimately in this example obtain all the roots of the problem. Indeed, this does raise the question of whether deflation should be able to obtain all roots of a nonlinear problem in this way starting from a single initial guess. This is a relevant question to which we will return in our 2-dof example that follows below.

\section{Phase Field Crystal (PFC) model for crystallisation of soft matter}

\begin{figure}[h]
\centering
{\vspace{-0cm}\includegraphics[width=7cm]{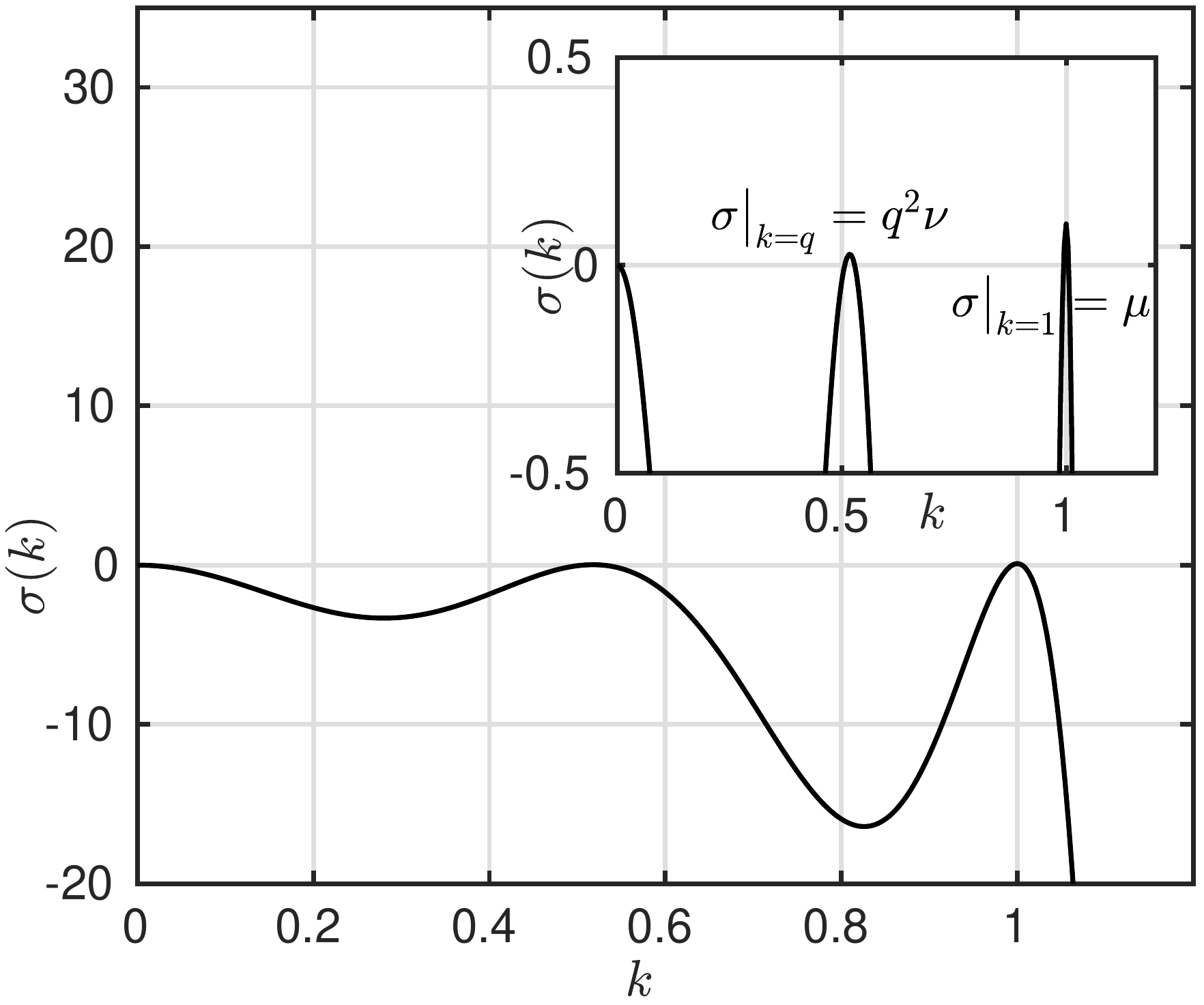}
}
\caption[]{Dispersion relation $\sigma(k)$ for the linear operator as a function of the wavelength $k$. Two peaks are visible at $k=q=1/2\cos(\pi/12)$ and at $k=1$. The height of the peaks in $\sigma(k)$ is related to the two growth rate parameters $\mu$ and $\nu$ as shown in the inset.}
\label{fig:disprel}
\end{figure}

The PFC model describes the density distribution of soft matter forming a liquid or solid on the microscopic scale and captures the associated time evolution over diffusive time scales. The scalar field $U(\textbf{x},t)$ describes the deviation of density at point $\textbf{x}$ and time $t$ about the average value. Resulting from the conservation of mass across the liquid/solid phase transition, the evolution is governed by a partial differential equation that evolves only to steady asymptotic states. The amount of useful work that can be extracted from this system being held at a constant volume and temperature, is the associated Helmholtz free energy ($\mathcal{V}$) in the system. In forming the free energy $\mathcal{V}$, we choose a linear operator ($\mathcal{L}$) and simple polynomial nonlinearities to include for nonlinear saturation (the $-U^4$ term) and to break the $U\rightarrow-U$ symmetry (the $U^3$ term). Then, the free energy $\mathcal{V}$ can be written in the form, 
\begin{equation}
\mathcal{V} = -\int \left[ \frac{1}{2} U\mathcal{L}U + \frac{Q}{3} U^3 - \frac{1}{4} U^4 \right] d\bx.
\label{eq:freeen}
\end{equation}
Here, the choice of $\mathcal{L}$ is done with the intention to promote two different length scales with wavenumbers, $k=q$ and $k=1$ independently. The choice of the linear operator $\mathcal{L}$ \cite{RucklidgePRL2012,SAKR16} allows for marginal instability at two wave numbers $k=1$ and $k=q$, with the growth rates of the two length scales determined by two {independent} parameters $\mu$ and $\nu$, respectively. The resulting growth rate $\sigma(k)$ of a mode with wave number~$k$ is given by a eighth-order polynomial:
 \begin{equation}
 \sigma(k)=\frac{k^2[\mu A(k)+\nu B(k)]}{q^4(1-q^2)^3}+\frac{\sigma_0}{q^4}(1-k^2)^2\,(q^2-k^2)^2\,,
 \label{eq:disprel}
 \end{equation} 
 where $A(k)=[k^2(q^2-3)-2q^2+4](q^2-k^2)^2 q^4$ and $B(k)=[k^2(3q^2-1)+2q^2-4q^4](1-k^2)^2$, which is also shown in Figure \ref{fig:disprel}. As shown in the inset, the peak height at the two chosen wave-lengths is related to the growth rates, which in this case are $\mu=\nu=0.1$. In the rest of this paper, we analyse this case for concreteness. Similar analyses can be
 performed for other choices of the parameters.

The relationship between the free energy $\mathcal{V}$ given above and the associated dynamics can be written in the form, 
\begin{equation}
\frac{\partial U}{\partial t} = -\nabla^2\left( \frac{\delta \mathcal{V}}{\delta U} \right) = -\nabla^2\left( \mathcal{L} U +Q U^2 -U^3 \right)\,.
\label{eq:pdedynamics}
\end{equation}
Here $U$ is a measure of density fluctuations around the mean density, the linear operator $\mathcal{L}$ chooses two different wavelengths for pattern formation as mentioned above and the parameter $Q$ quantifies the effect of destabilising second order nonlinearities.

This system is a good example to use in this work as previous results indicate that the solution landscape of this model is rich with many different types of patterns. The dynamics conserves the mass (i.e., the integral of $U$ over all of space) in the system. This implies that the dynamics will only possess equilibria that are steady in time. Such a solution landscape which does not have (propagating) waves or other time varying equilibria simplifies the analysis of applying different techniques to determine equilibria. Another advantage is that the same model also can be used to analyze crystallisation of soft matter in 3D. Hence, it bears a number of advantages in
connection to our goals herein and provides a ripe testbed also for possible extensions.

\subsection{Reduction to coupled ODEs}
\label{sec:2dsys}
  
\begin{figure}[h]
\centering
{\vspace{0.3cm}\includegraphics[width=5.5cm]{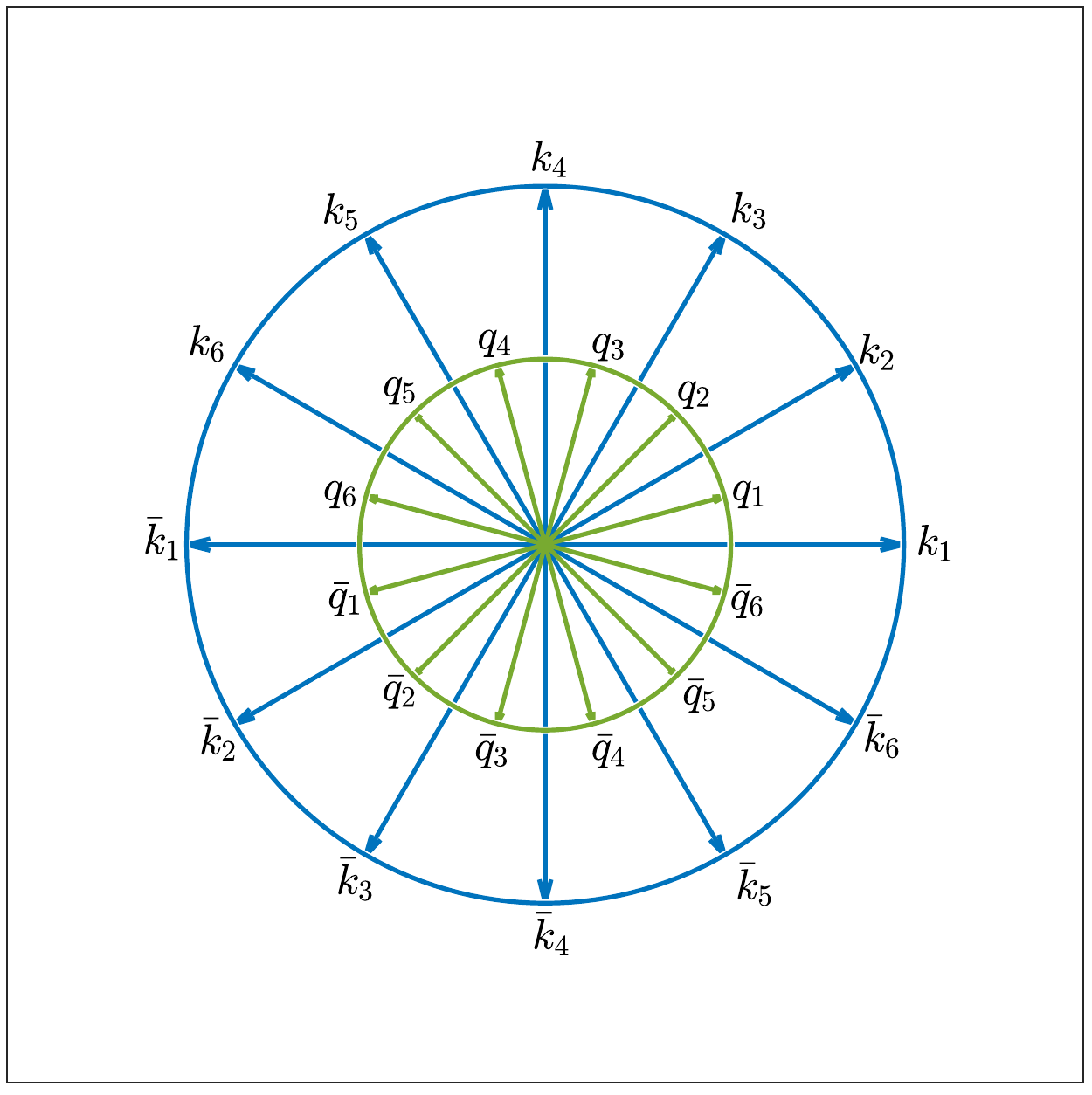}
\hspace{0.7cm}
\includegraphics[width=6cm]{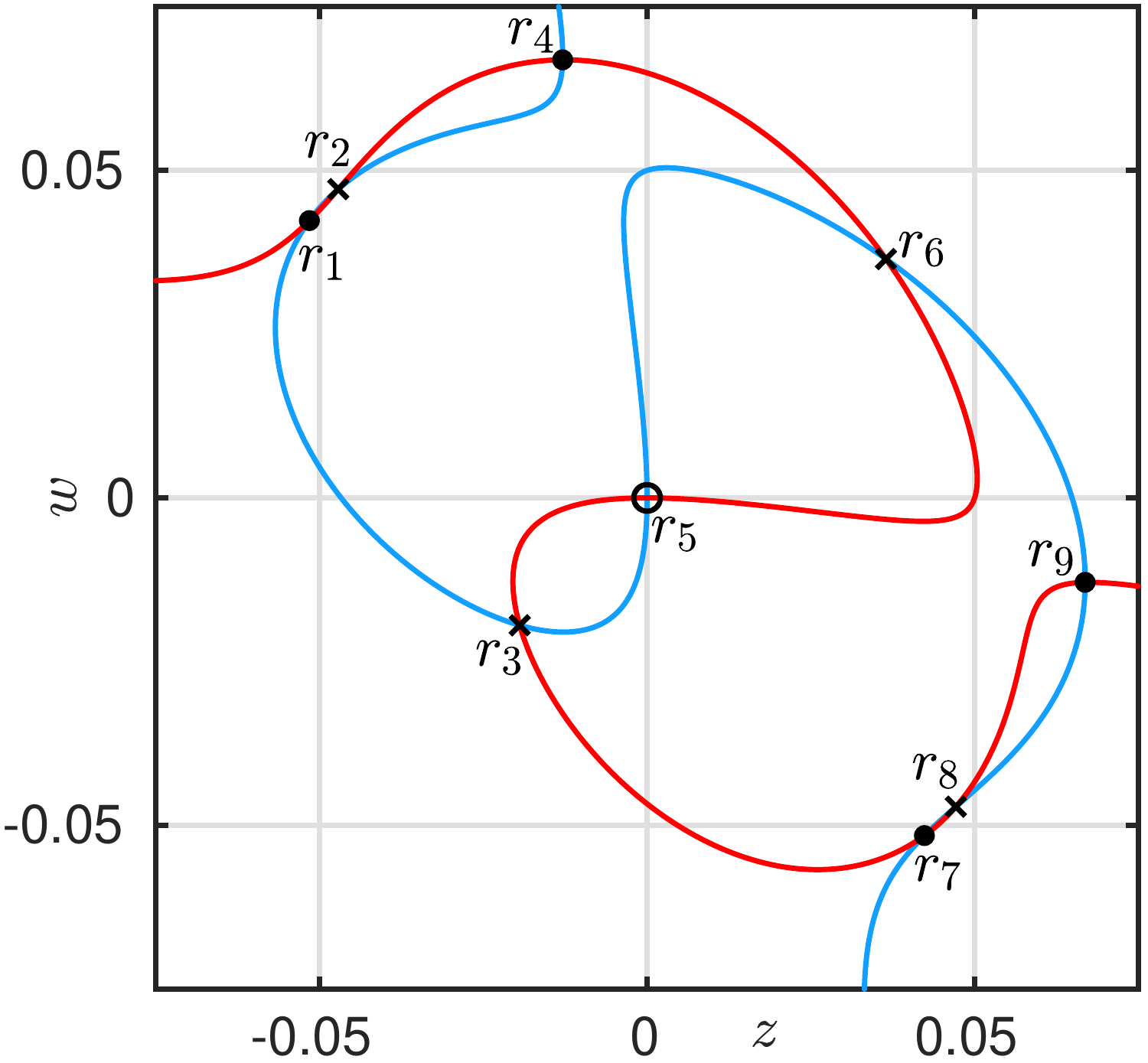}
}
\caption[]{Left: Wavevectors at two lengthscales, $|k|=q=0.5176$ and at $|k|=1$ that form a dodecagonal quasicrystal. Wavevectors $k_j$ have modulus $|k|=1$ and $q_j$ have  $|k|=|q|$ for $j=1, \dots,6$.  
    Complex conjugate wave vectors are identfied with bar, i.e., $\bar{k}_1=-k_1$. Right: Solution curves for the two coupled polynomials in (\ref{eq:ampeqn}) with the zero contour of
  the right hand side of the first equation in~(\ref{eq:ampeqn})
  in blue and that of the second equation in red. Solutions of both polynomials indicated with black markers: hollow circles for maxima, crosses for saddles and filled dots for minima.}
\label{fig:12foldfft}
\end{figure}

  As explained in the previous section, the growth rates $\mu$ and $\nu$ dictate the height of the dispersion peak at the two chosen length scales $k=q$ and $k=1$. At the onset of pattern formation, where both  length scales are promoted, we have the situation with both $\mu$ and $\nu$ being marginal at a co-dimension 2 bifurcation. 
  Close to this onset, we can assume that the resulting pattern in the density field $U$ is formed of wave-vectors at these two marginal lengthscales. This then allows us to perform a weakly nonlinear analysis by expanding the density field $U$, the linear operator model parameters $\mu$,$\nu$, the quadratic nonlinearity prefactor $Q$ and time $t$ in terms of a small parameter $\epsilon$ via a multiple scales expansion.
For the choice of scalings with $U=\epsilon U_1$, $\mu=\epsilon^2\mu_1$, $\nu=\epsilon^2 \nu_1$, $Q=\epsilon Q_1$ and $t=\epsilon^{-2}t$, we are able to balance the effects of the parameters, nonlinearities and time at the same order of expansion. Additional details regarding the selection of the scales of the expansion
can be found in~\cite{SAKR16}. We further write $U_1$ in terms of a finite number of marginal modes as 
\begin{equation}
  U  = \sum_{j=1}^{N} \left( z_j e^{i\textbf{k}.\textbf{x}} + w_j e^{i\textbf{q}.\textbf{x}}\right) + c.c.
  \label{recon}
\end{equation}
For the choice of $q=(1/2) \cos(\pi/12)\approx 0.5176$, we promote the formation of 2D dodecagonal quasicrystals, when the number of marginal modes is 12 in each wavelength, i.e., $N=6$. The left panel of Figure \ref{fig:12foldfft} shows a representative example of the wave vectors involved in creating a 12-fold quasicrystal.

Substituting the above expression into Eq. (\ref{eq:freeen}), we can write the volume specific free energy $v={\mathcal{V}}/({\epsilon^4\,\textrm{Volume}})$ as
\begin{eqnarray}
v = -\mu z_1 \bar{z}_1 &-& 2Q\left( z_3 \bar{z}_5 + w_1 \bar{w}_6 + w_3 \bar{z}_6 + \bar{w}_4 z_2 \right) \bar{z}_1 -\mu \sum_{j=2}^{j=6} \mu z_j \bar{z}_j - \sum_{j=1}^{6} \nu w_j \bar{w}_j\nonumber\\
&-& Q \textrm{(28 other cubic terms)} - \textrm{(148 quartic terms)}\,.
\end{eqnarray}
where (similar to~\cite{SAKR16}) we have written the contributions involving $\bar{z}_1$ explicitly up to third order.
Assuming full symmetry among all wave vectors with the same wave number, i.e., $z_j=z$ and $w_j=w$ for all $j$, this expression becomes 
\begin{equation}
v = \left(-\mu z^2 -\nu w^2 - \frac{4Q}{3}w^3 -4Q w^2 z - 4Q w z^2 -\frac{4Q}{3} z^3 + \frac{99}{6} w^4 + 24 w^3 z + 60 w^2 z^2 + 24 w z^3 + \frac{99}{6} z^4\right)\,.
\end{equation}
Taking functional derivatives with respect to $z$ and $w$ then gives us the coupled amplitude equations for the evolution of $z$ and $w$ as a gradient system
associated with this free energy:
\begin{eqnarray}
\dot{z} =-\frac{dv}{dz}&=& 2\left( \mu z + 2 Q z^2 + 4 Q w z + 2 Q w^2 - 12 w^3 - 60 w^2 z - 36 w z^2 - 33 z^3\right)\,,\nonumber \\
\dot{w} =-q^2 \frac{dv}{dw}&=& 2 q^2\left( \nu w + 2 Q w^2 + 4 Q w z + 2 Q z^2 - 12 z^3 - 60 w z^2 - 36 w^2 z - 33 w^3\right)\,.
\label{eq:ampeqn}
\end{eqnarray}

\subsection{ODE Equilibria}

In order to obtain the equilibria of the above coupled equations, we convert them to two coupled polynomials, 
\begin{eqnarray}
f_1 &=& \mu z + 2 Q z^2 + 4 Q w z + 2 Q w^2 - 12 w^3 - 60 w^2 z - 36 w z^2 - 33 z^3\,, \nonumber \\
f_2 &=& q^2 \left(\nu w + 2 Q w^2 + 4 Q w z + 2 Q z^2 - 12 z^3 - 60 w z^2 - 36 w^2 z - 33 w^3\right) \,.
\label{eq:polysys}
\end{eqnarray}
Here, our emphasis is
on the proof of principle of the techniques rather than on an exhaustive parametric
study. Therefore, we solve this set of equations for a given value of $\mu=\nu=0.1$ and $Q=0.3$ and do not perform a detailed parametric study. Similar considerations to the ones discussed below for this set of parameters can
naturally be extended to other parametric combinations. We first solve the above coupled polynomials using homotopy methods (Bertini)~\cite{bertini}. Given the features of this numerical algebraic geometry
software~\cite{bertini}, we are guaranteed to obtain all real equilibria of the system. The right panel in Figure \ref{fig:12foldfft} shows the zero contours of the two functions $f_1$ (in red) and $f_2$ (in blue).
It is worth mentioning in passing that these zero contours are central
to suitable methodologies for identifying roots (that we do not focus
on here) such as the reduced
gradient following method~\cite{quapp}. 
Their crossing points then identify the nine possible equilibria (marked in terms of $r_1$ to $r_9$ in the right panel of Figure \ref{fig:12foldfft} and later). By determining the eigenvalues of the linear operator at each solution, we can determine the dynamical stability of each of the determined equilibria in the context of the
reduced ODE model of Eqs.~(\ref{eq:ampeqn}). We indicate the stability information of each equilibrium in the figures with black filled dot indicating a stable minimum, black circle indicating an unstable maximum and black cross indicating a saddle point. The interpretation of these eigenvalues within the realm of the original PDE problem will be briefly discussed in the next section.
 
\begin{figure}[]
\centering
{
\includegraphics[width=6cm]{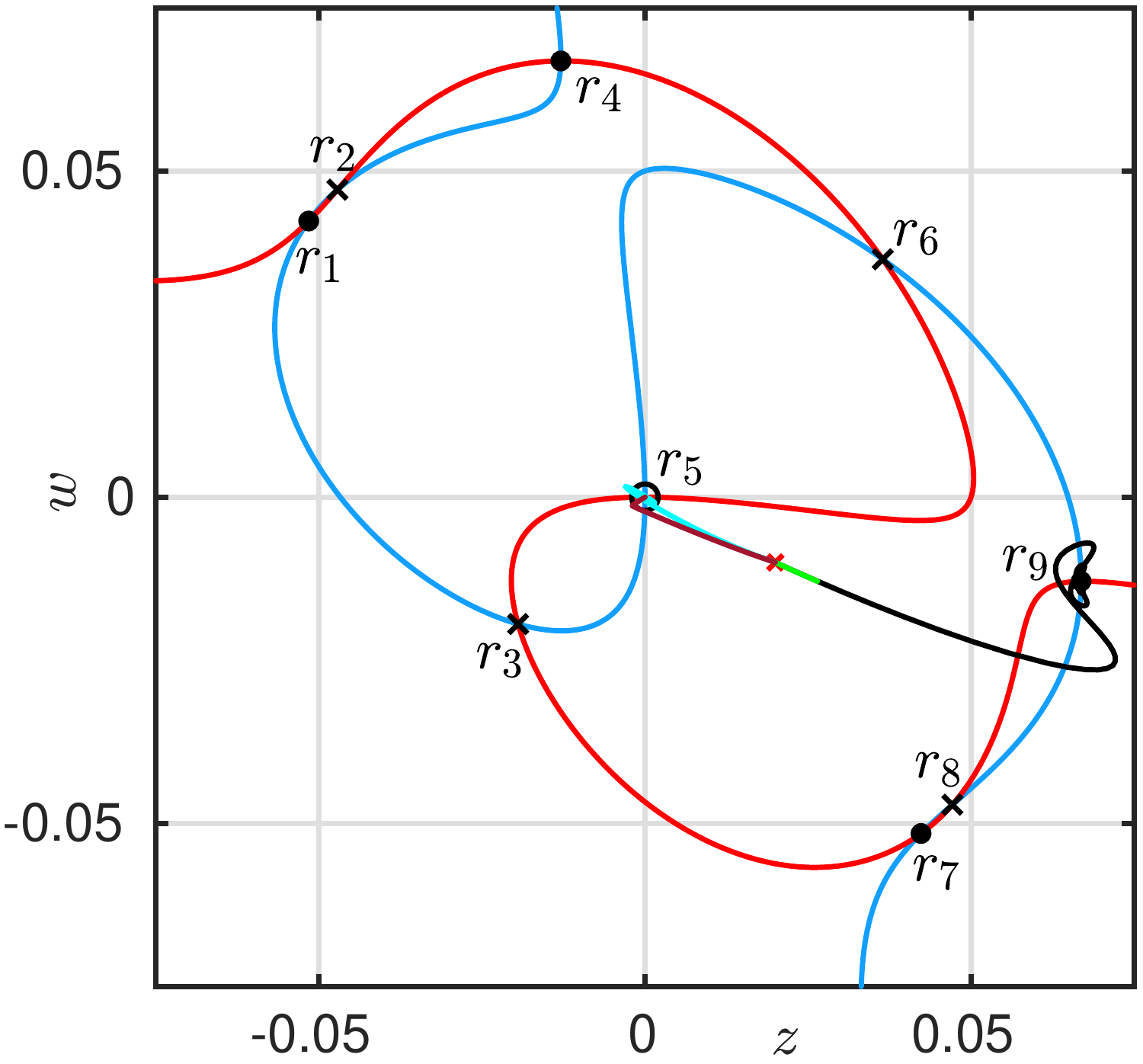}
\hspace{1cm}
\includegraphics[width=6cm]{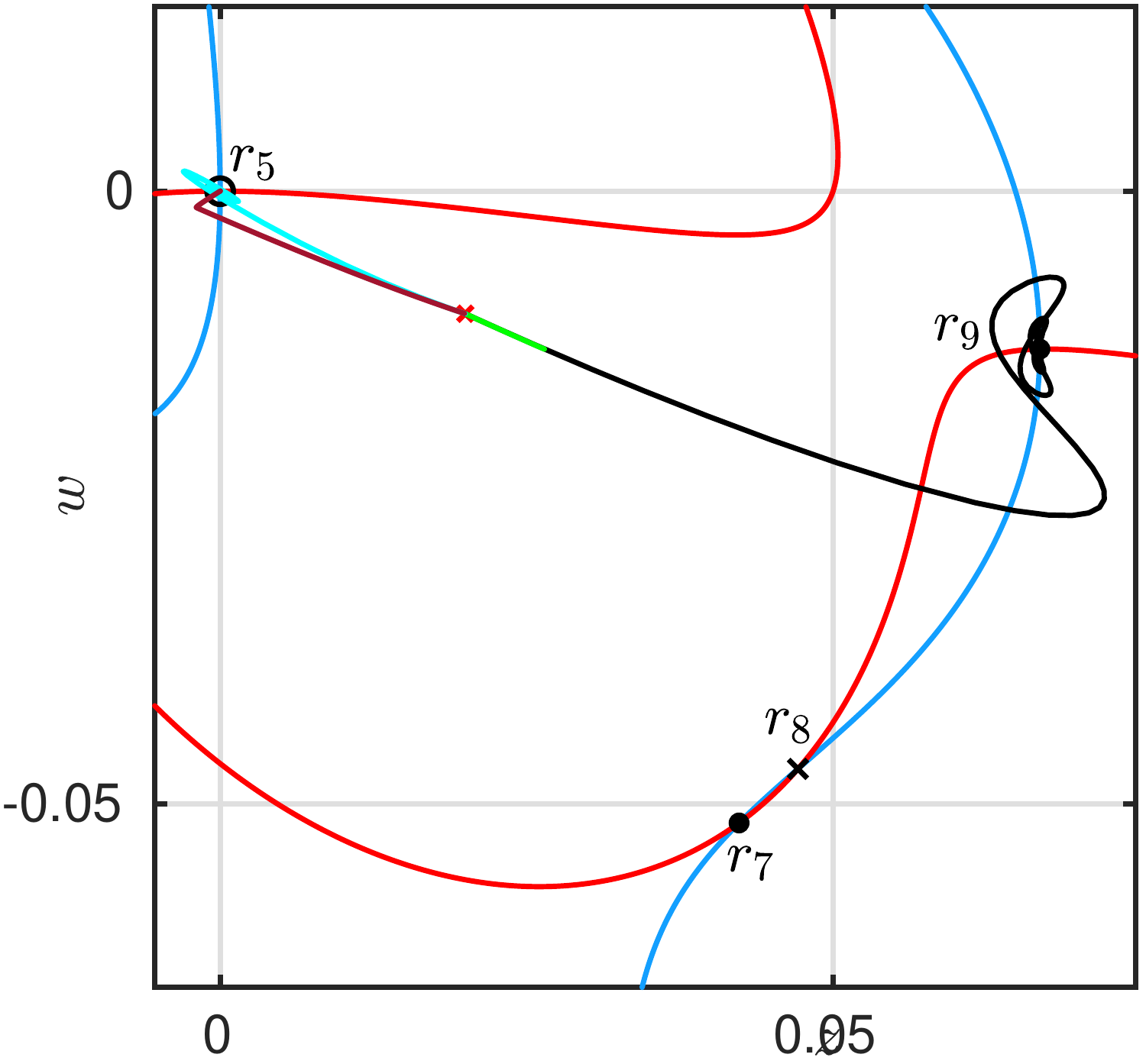}
}
\caption[]{Left: Evolutions using CTN (in black) and CTN with deflation (green). Evolutions using SOM combined with CTN but without (cyan)  or with (brown)
  deflation. Right: Zoomed in view of left panel. We see that both CTN and SOM behave similarly with and without deflation. The evolution after deflation in both cases still gets trapped to the original equilibrium which is now a pole. The system parameters are $\mu=\nu=0.1$ and $Q=0.3$.}
\label{fig:2DFlowDeflation}
\end{figure}

If the system of interest is $f (z,w)= [f_1,f_2]^T$, in addition to exploring
CTN methods for $f$ itself, we can also define a related squared operator as in the 1D example as $g(z,w)$, which will have as its minima, all of the equilibria of $f(z,w)$. The first step to try is to employ the combination of time stepping in terms of CTN on $f(z,w)$ or $g(z,w)$. Results of this are shown in Figure \ref{fig:2DFlowDeflation}. In the left panel, we see the solution landscape of the system $f$ just as in the right panel of \ref{fig:12foldfft}. The right panel of Figure \ref{fig:2DFlowDeflation} shows a zoomed in view of the left panel. The overlay of blue filled circles at every equilibrium indicates that in the associated squared operator system of $g(z,w)$, all the original equilibria are now minima. Moreover, this panel shows that there are no additional minima that have been created in the transformation. {While it is guaranteed in this case, by construction of the effective landscape (associated
with $f^2$) that {\it only} the roots of the original system will constitute global minima of the associated squared-operator system, in principle additional nonzero minima, as well as separatrices between them, may arise in general.}
Starting from the initial conditions of $(z,w)=(0.01,-0.01)$ (indicated with a red cross), CTN on $f$ shown as black curve approaches the minimum denoted as $r_9$. Starting from the same initial condition, CTN on $g$ (shown as cyan curve) approaches the much closer equilibrium of $r_5$ which, recall, is a minimum for the SOM, but
an energy maximum for the original system. This is naturally in line with energetic considerations in these flow-based systems.

The next step that we considered (in analogy with our 1-dof exposition) was to try to include deflation in conjunction with the CTN method. When we try to deflate $r_9$ for the CTN on $f$, the new dynamics (shown as green curve) seems to follow the un-deflated black curve (i.e., a similar trajectory as before) before stopping. This occurs due to the fact that in the deflated dynamics, the earlier energy minimum has become a pole where the particle can slide
towards indefinitely negative effective potential energy. While this landscape interpretation may no longer be formally possible (as the system may not be, by construction, a gradient one), nevertheless, it is used here by analogy with the 1-dof case (where such an interpretation can always be brought to bear). The numerical time stepping algorithm accordingly adjusts the step-size and takes smaller and smaller steps till the rate of increase exceeds the limits set by the smallest allowed stepsize and the algorithm stops. Nevertheless, even if it did continue, the above analysis suggests that it would not reach the desired result (i.e., an alternative root). A similar problem persists for the CTN on $g$ (i.e., combined with SOM). After deflation of $r_5$ as well, the system follows the brown curve, which is a nearby curve to the one without deflation (cyan). Hence we see that the pole problem that we encountered in the 1D example, when combining flow methods with deflation, persists, as may be natural to expect.

\begin{figure}[]
\centering
{
\includegraphics[width=7cm]{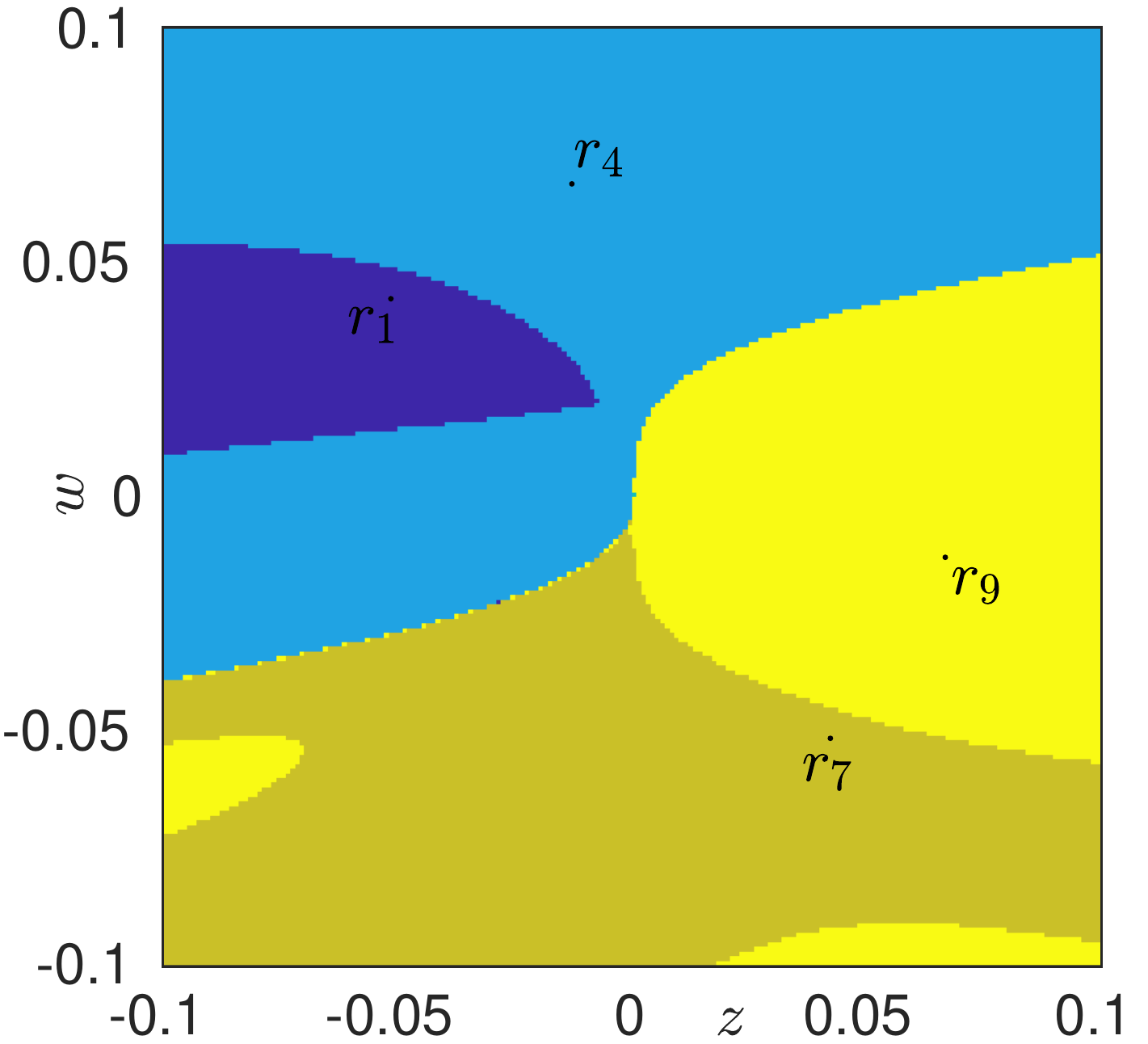}
\hspace{0.7cm}
\includegraphics[width=7cm]{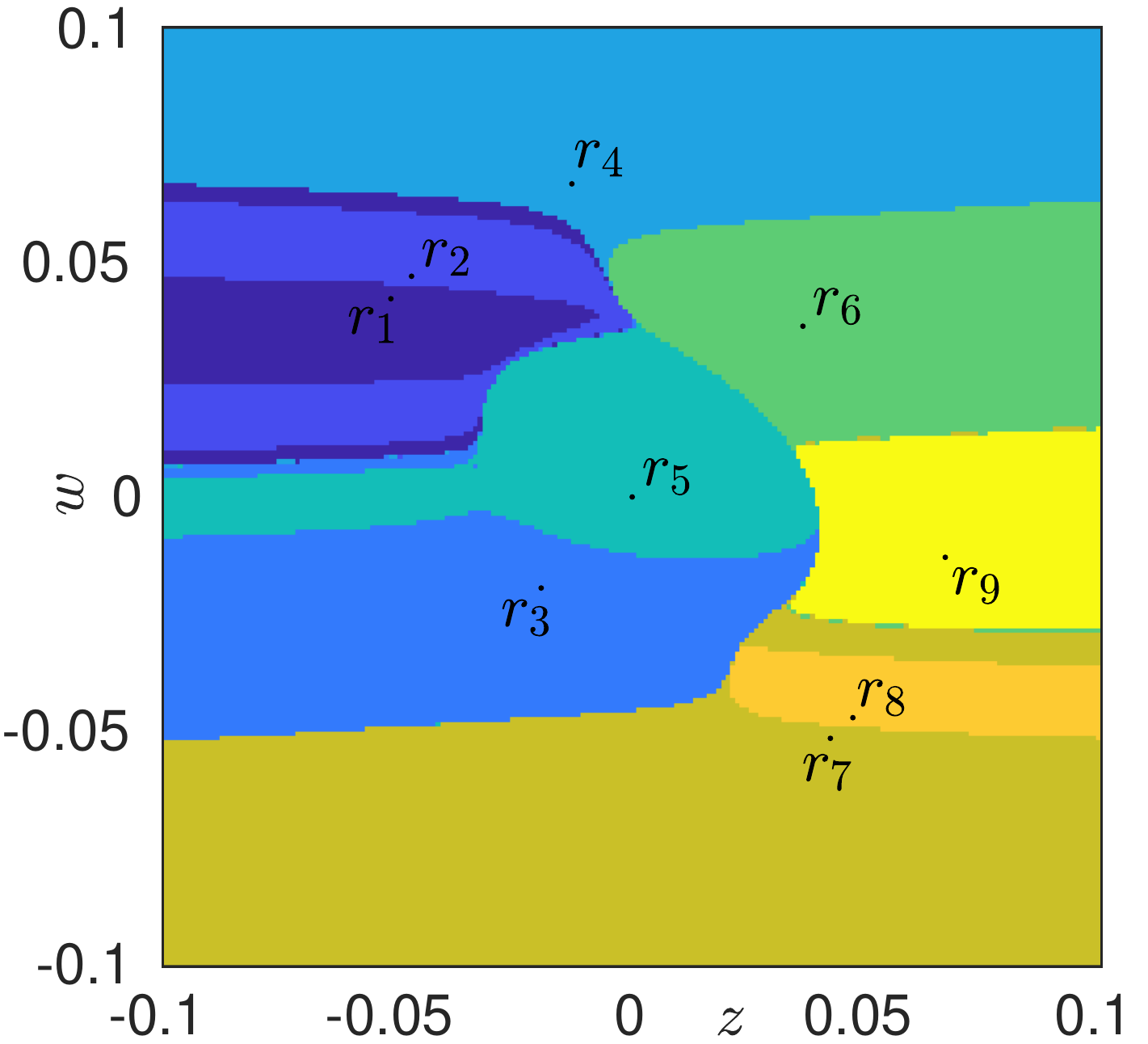}
}
\caption[]{Left: Asymptotic results of the dynamics stemming from
  evolving the
  CTN system with initial conditions on the $(z,w)$ plane and
  vanishing initial speeds. The space of possible initial conditions in the
  $(z,w)$ plane has been split into four regions, each corresponding to one of the 4 stable equilibria of the original CTN system. The stable equilibria are shown as black filled circles for reference. Right: the same view but now from combining SOM with CTN. Here we observe the non-trivial partitioning of initial conditions in the plane of $(z,w)$ evolving to one of the 9 equilibria of the original CTN system. Stable equilibria of the SOM-CTN system shown as black filled circles (with the same naming convention as in Fig. \ref{fig:12foldfft}($b$) and later). System parameters are $\mu=\nu=0.1$ and $Q=0.3$.}
\label{fig:2Dbasin}
\end{figure}   
Having encountered problems upon starting from a single initial
condition and using deflation combined with flow methods to recover
all equilibria in this system, we try to use flow methods starting at
different initial conditions to identify the basins of attraction of
each minimum in the system. By the term `basins of attraction', in the case of the present non-autonomous systems,
we will mean the regions leading to the same asymptotic outcome of the dynamics starting from arbitrary initial conditions within the plane of $(z,w)$, i.e., within the chosen symmetry subspace, along with vanishing initial speeds.
Results of this analysis are in Figure \ref{fig:2Dbasin}. The left panel shows the basins of attraction of the minima for the 2-dof vector field $f$ where a point is colored depending on the final equilibria that a flow method starting from that point would converge to. Minima of the system are marked with black crosses. It is interesting to compare this figure with the right panel of Figure~\ref{fig:12foldfft}. The latter encompasses the saddle points of the energy landscape, while in
the left panel of  Figure~\ref{fig:2Dbasin} it is only possible to converge to the energy minima.

The right panel in Figure \ref{fig:2Dbasin} shows the basins of attraction for the squared operator system $g(z,w)$ along with the minima for $g$ shown as black crosses. This result shows us that using flow methods for the squared operator dynamics allows to obtain all the equilibria of $f$, upon suitable scanning of the two-dimensional space of associated initial conditions of the SOM flow method. Naturally, there is a region containing the immediate neighborhood of each (global) minimum that results in convergence to it. However, both figures appear to be somewhat more complex in that there exist regions of one color within a domain of the two-dimensional space that is predominantly of a different color. This presumably has to do with the presence of some of the unstable nodes
and saddle points emerging in the modified energy landscape $V_{eff}=(1/2) (f_1^2+f_2^2)$, as well as with the effectively $4$-dimensional nature
of the CTN systems (due to its involving accelerations in each of the
two dynamical variables $(z,w)$).
  
\begin{figure}[]
\centering
{\vspace{0.3cm}\includegraphics[width=8cm]{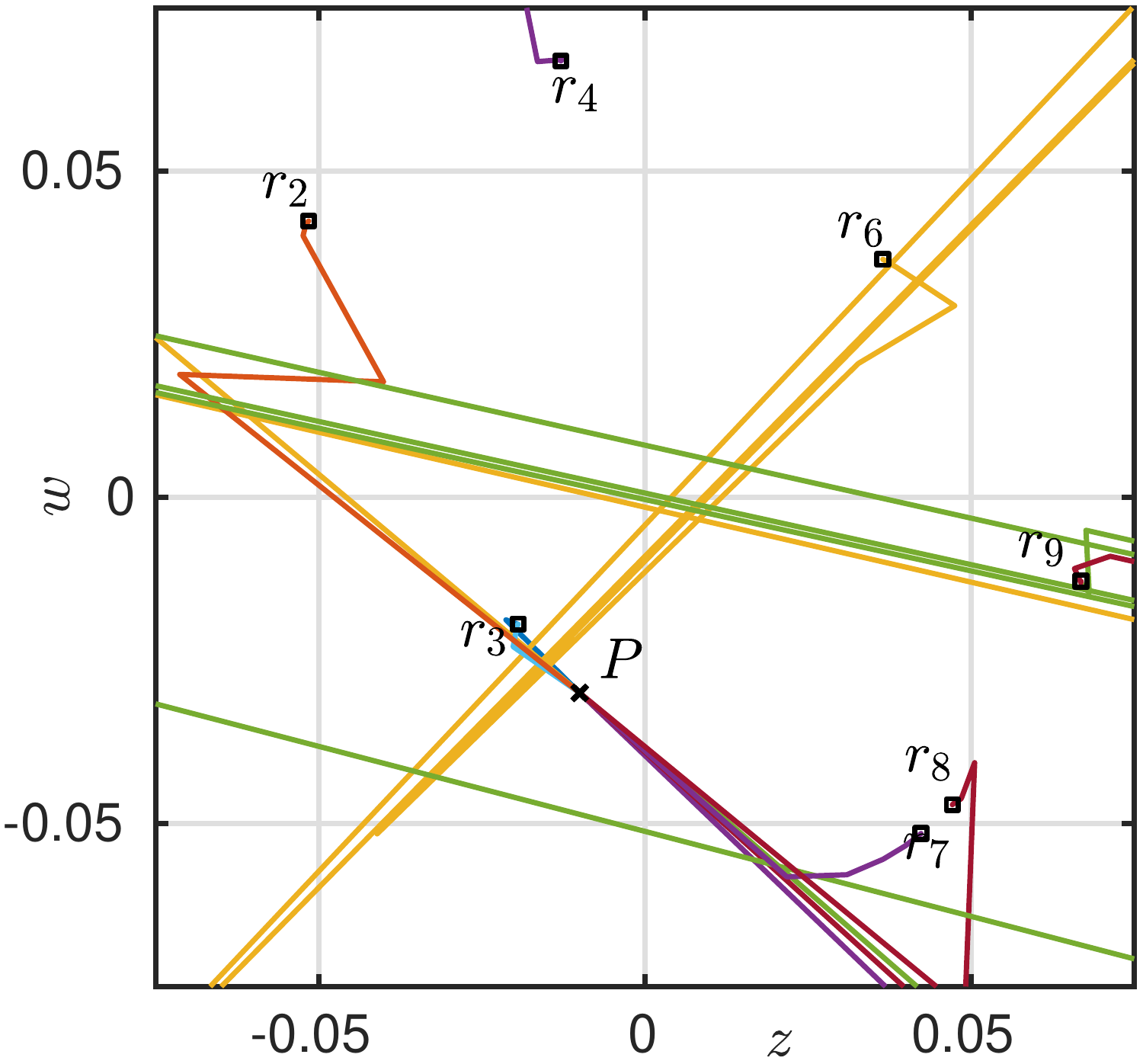}}
\caption[]{Newton with deflation for the 2D system plotted as a function of the two unknown amplitudes $z$ and $w$. Starting from the same initial guess at $(z,w)=(-0.02,-0.03)$ (shown as a black cross) and sucessively using deflation, we are able to converge to 7 of the 9 equilibria in this system (shown as black squares).
}
\label{fig:2DNewtonDeflation}
\end{figure}  
  
Next we obtain the equilibria of the system $f(z,w)$ using Newton's method coupled to deflation in the 2-dof system. The results of this analysis are summarised in Fig.~\ref{fig:2DNewtonDeflation}. Choosing $(z,w)=(-0.01,-0.03)$, shown as black cross as the initial guess, the Newton iteration first converges to the equilibrium $r_3$ at $(-0.0194,-0.0194)$. When this solution is deflated, the Newton iterations in Fig.~\ref{fig:2DNewtonDeflation} follow the purple line and converge to $r_4$ at $(-0.0129,0.0669)$. Deflating both of these solutions, we see that successive Newton iterations follow the dark red line in Fig.~\ref{fig:2DNewtonDeflation} and converge to the saddle at $r_8$ at $(0.0471,-0.0471)$. Further deflations (in different combinations) allow us to converge to the additional equilibria $r_2$, $r_6$, $r_7$ and $r_9$. We see that in this case, the additional equilibria $r_1$ and $r_5$ were not approachable using Newton iteration with deflations of previously determined equilibria. However, choosing other initial conditions, e.g., $(-0.01,-0.01)$, will allow us to uncover these two equilibria easily. An interesting conclusion from these still relatively straightforward/tractable, yet quite instructive considerations is that, in line with what was also found, e.g., in~\cite{egc_16,egc_20} in an infinite-dimensional (and hence much richer) problem, there are no a priori guarantees that deflation from a given initial guess will converge to all solutions. Indeed, even in our simple 2-dof setup, the method  is not able to converge to all solutions. Hence, it is relevant to seek a broader representation of the space of initial guesses in order to capture the widest possible solution span.
 
 
\subsection{Extension to the Original PFC Problem}

The next question to ask is how much of the information from the ODE system remains relevant to the dynamics of the (considered numerical
discretization of the) full PDE system. In addition to seeking to identify the relevant configurations at the level of the PDE, we will also seek to connect the ``reduced stability'' of the 2-dof system with the full PDE stability at the level of the original PFC system bearing infinite degrees of freedom. The two main assumptions that allow for the reduction from a PDE to an ODE system as derived in section \ref{sec:2dsys} are, ($i$) the dynamics is assumed to be in the neighbourhood of a critical point, here a co-dimension 2 bifurcation and ($ii$) a specific symmetry is assumed in the form of marginal modes, here wave vectors corresponding to a dodecagonal quasicrystal which  are chosen in Eqn.\,(\ref{recon}). This implies that the predictions from the ODE system should match the PDE dynamics close to the co-dimension 2 bifurcation when the PDE dynamics is restricted to the dodecagonal symmetry subspace. 

With respect to relaxing the first assumption, previous work such as \cite{Wittenberg1997} has pointed out that the occurrence of other global bifurcations close to the critical point of interest can restrict the region in parameter space where the ODE system reproduces the dynamics of the PDE system. However, in a generic case with no other global bifurcations in the neighbourhood of the critical point in phase space, the predictions of the reduced ODE system can match well  the PDE
over a wide range of parameters around the critical point. We discuss how the current system falls in this generic category by determining related PDE equilibria to each determined ODE equilbrium. 

\begin{figure}[]
\centering
{\includegraphics[width=7.8cm]{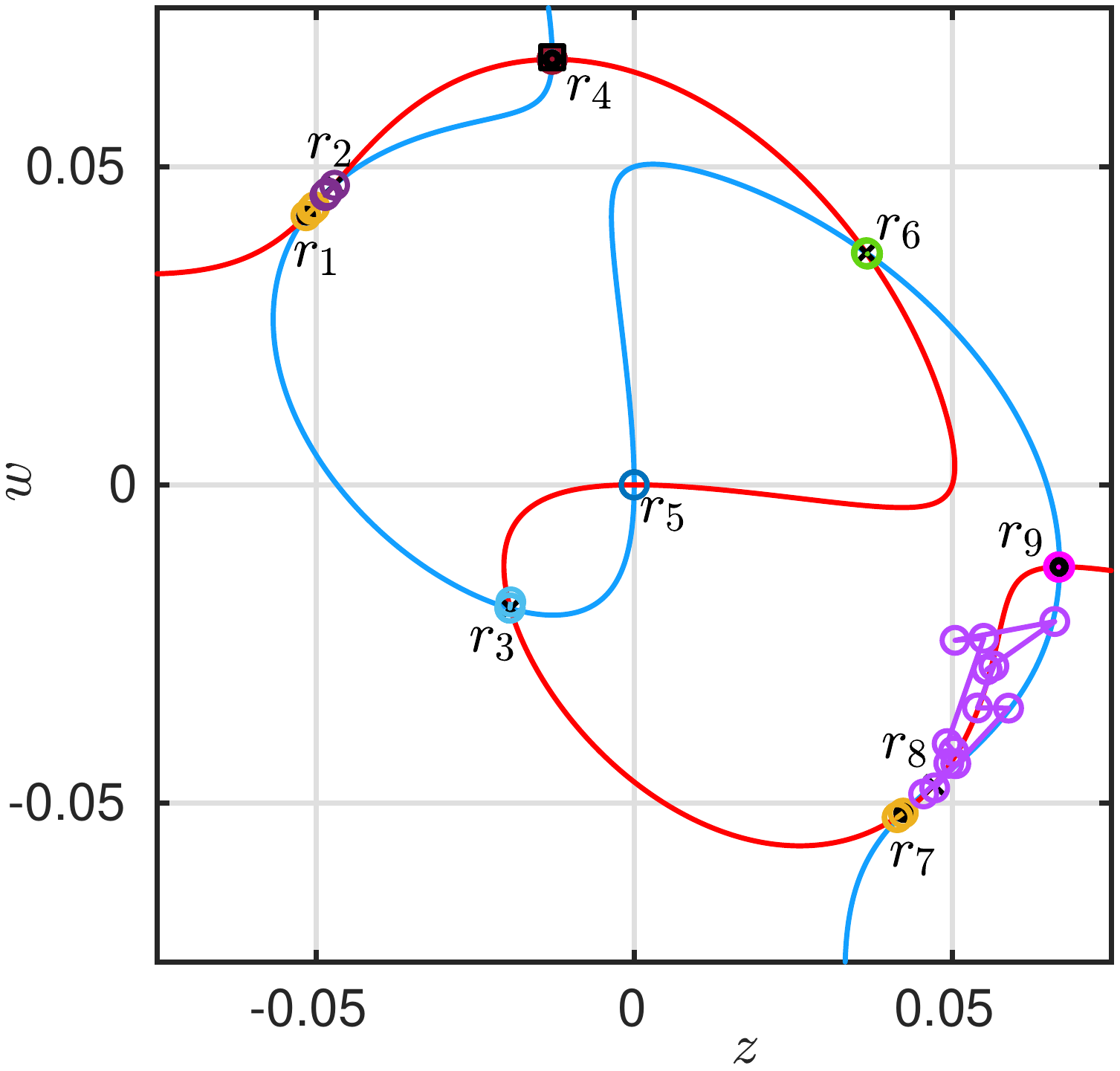} }
\caption[]{Comparison of Newton iteration for the PDE system with equilibria of the ODE system as initial guess. Each Newton iteration is shown as a colored circle. The black square is the equilibrium of the PDE system related to $r_4$ of the ODE system, which is discussed in more detail in Figure \ref{fig:2DODEPDE}. System parameters are $\mu=\nu=0.1$ and $Q=0.3$. 
}
\label{fig:2DODEPDENewton}
\end{figure}

Firstly, we deploy a Newton method to check the existence of equilibria for the full PDE system corresponding to the ODE solutions. In this process, it is relevant to bear in mind that we can recover associated equilibria for the PDE even if they are dynamically unstable. The results of our Newton iterations are shown in Figure \ref{fig:2DODEPDENewton} with colored circles indicating converged results at each Newton iteration starting from different ODE equilibria are indicated with differently colored circles. The observation is that for the case with $\mu=0.1$ and $\nu=0.1$, all of the ODE solutions can be recovered as solutions to the full PDE system using iterative methods for the full PDE. The reconstruction of the corresponding ODE solutions at the level of PDE waveforms via Eq.~(\ref{recon}) allows to construct suitable, rapidly converging (at the level of the Newton iteration) initial guesses.  

\begin{figure}[]
\centering
{\includegraphics[width=15cm]{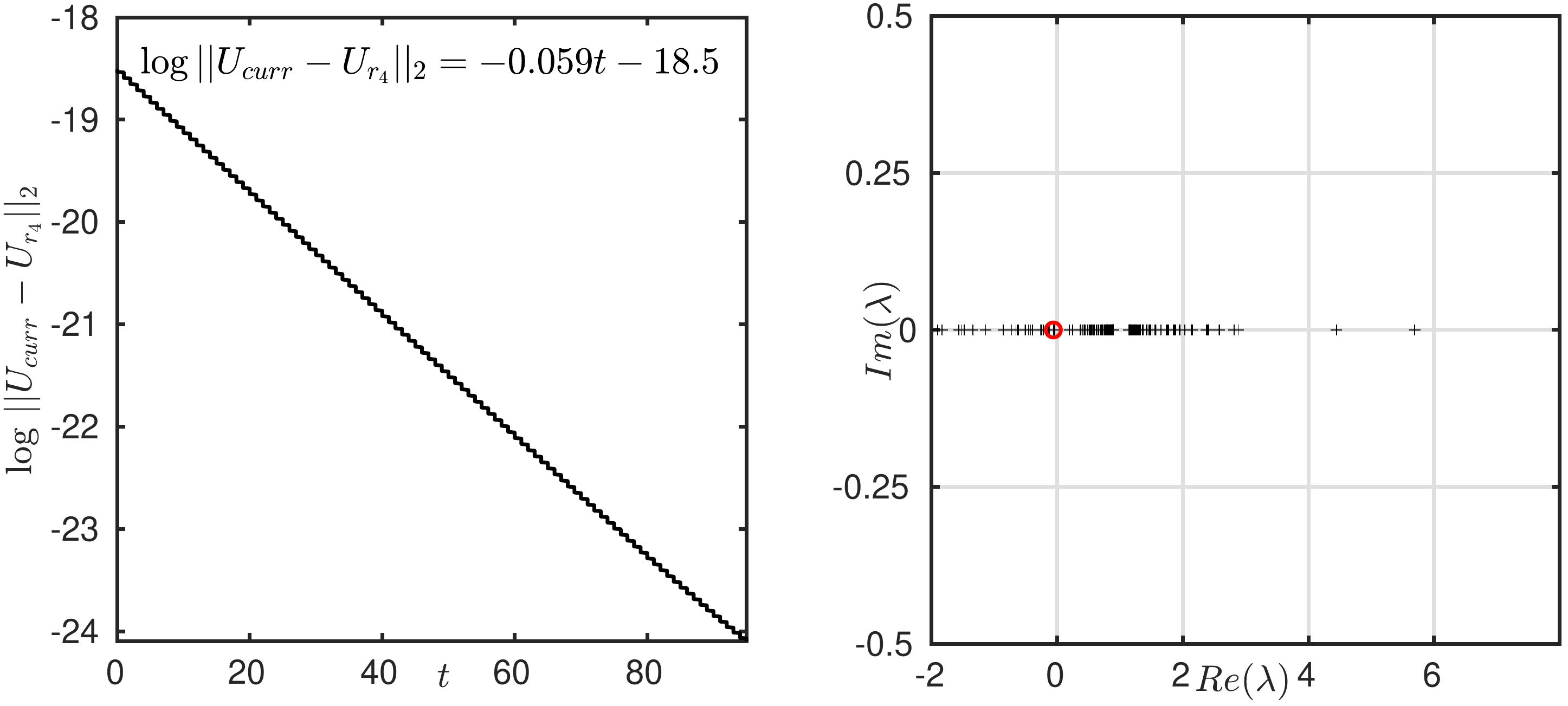}
}
\caption[]{
Left panel: Evolution of a perturbation along the eigenvector corresponding to the largest eigenvalue $\lambda_1=-0.061$ close to the equilibrium $r_4$. Slope of this evolution in log scale $-0.059$, closely matches the predicted rate of decay for the largest eigenvalue $\lambda_1$, Right panel: Comparison of eigenvalues from the ODE (in red circles) and corresponding PDE solution (in black crosses) for $r_4$. System parameters are $\mu=\nu=0.1$ and $Q=0.3$. 
}
\label{fig:2DODEPDE}
\end{figure}

Secondly, we compare the stability information for the ODE equilibria with an estimate for their stability in the PDE system from evolutions of the PDE system. Let us consider the example of the ODE equilibria at $r_4$ which is a minimum with eigenvalues: $\lambda_1=-0.061$ and $\lambda_2=-0.058$. The eigenvector corresponding to the largest eigenvalue, $\lambda_1$ is along the vector $(z,w) =(-0.4479,1)$. We construct an initial condition for the PDE system by multiplying this eigenvector with a small perturbation amplitude, e.g., $10^{-11}$ and adding it to the state $r_4$. During evolution of the PDE system from this initial condition, we observe that the perturbation decays and we recover the equilibrium $r_4$ as the asymptotic state. We track the decay of the perturbation as a function of time in log scale as shown in the left panel of  Fig.\,\ref{fig:2DODEPDE}. We see that in this case, the slope of the decay in log scale is $-0.059$, which is close to the value predicted as the decay rate $\lambda_1$. Similarly we also confirm that positive eigenvalues predicted from the ODE system match the linear growth rate of perturbations along the corresponding eigenvector. For example, we consider the saddle at $r_2$ with positive eigenvalue $\lambda_1=0.005758$. Perturbation along the corresponding eigenvector in the PDE system grows at the rate of $0.005294$ (not shown). Both the above comparisons of ODE eigenvalues along with the decay/growth rates observed in the PDE evolutions started with the eigenvectors as initial perturbations shows us that the equilibria of the reduced ODE system can persist in the full PDE system with the same symmetry as long as the initial perturbations are within the dodecagonal symmetry subspace spanned by the marginal modes in Eqn.\,(\ref{recon}). 

The second assumption in the reduction to an ODE system is that of a choice of symmetry subspace. Given that the PDE system is not constrained to evolve within the chosen subspace, we check the stability of the obtained PDE equilibria against all possible initial perturbations. We do this by determining the leading eigenvalues of an equilibria using the matlab `eigs' subroutine with the option of `bothendsreal' which ensures that we obtain the eigenvalue with the largest positive real part as $\lambda_{PDE}=5.69$. The eigenvalues calculated for the PDE equilibrium related to $r_4$ are shown in the right panel of Figure \ref{fig:2DODEPDE}. We note that there are corresponding eigenvalues in the PDE solution at the eigenvalues predicted by the ODE solution, however we see that the PDE solution is highly unstable with multiple positive real eigenvalues (shown as black crosses). Therefore, we expect flow methods for the PDE to not recover the ODE solution in the asymptotic limit for generic initial perturbations that do not reinforce any symmetry. 

\begin{figure}[]
\centering
{
\includegraphics[height=6cm]{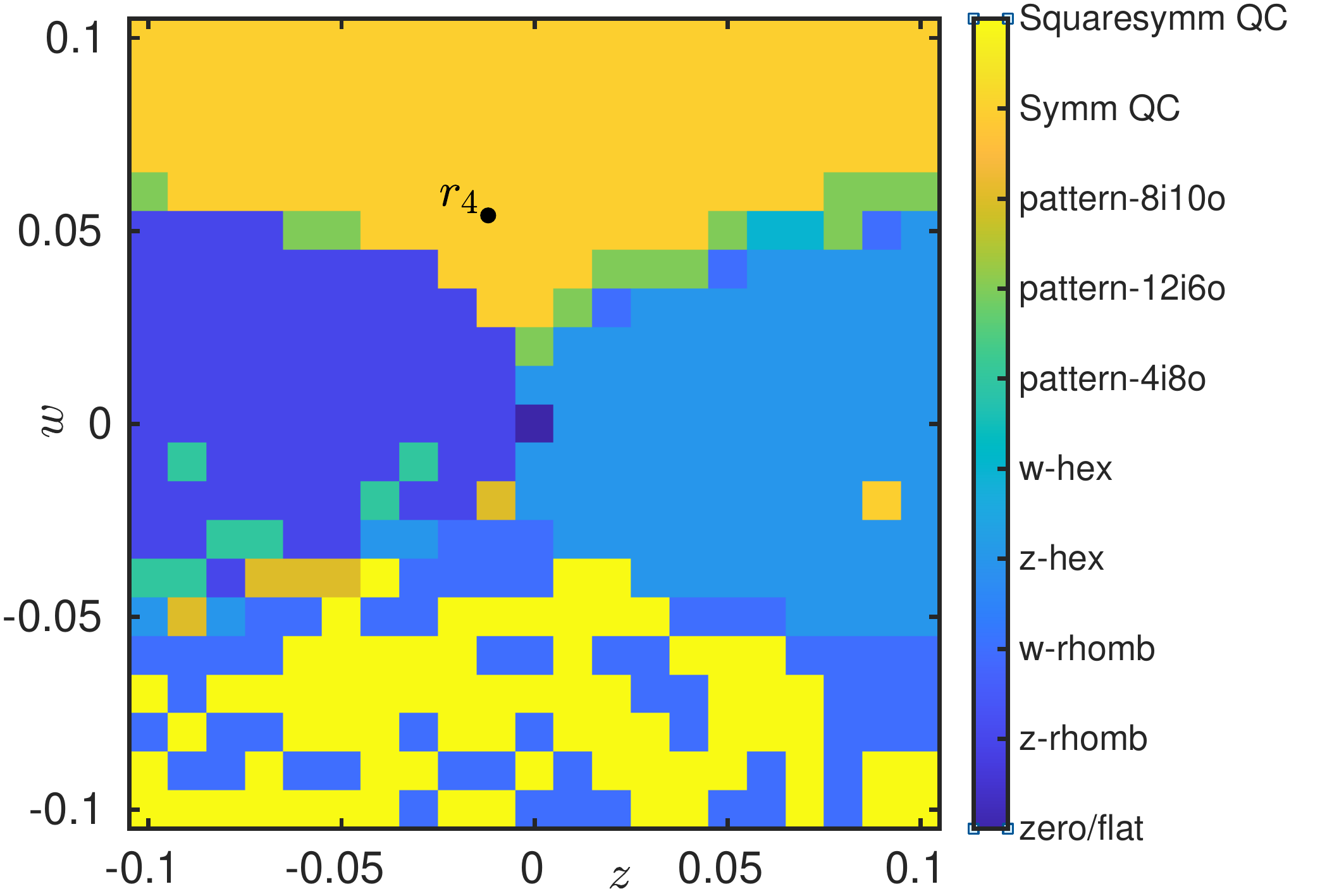}
}
\caption[]{Dynamical outcomes of different initial conditions
  reconstructed
  through the projection in the $(z,w)$ plane for the
  PDE system
  yielding different types of patterns.}
\label{fig:2Dpdebasin}
\end{figure}

In the next step, we examine the asymptotic results of the PDE dynamics when we consider initial conditions that fall within the dodecagonal symmetry subspace with different values for $(z,w)$ similar to the investigation in the left panel of Fig.\,\ref{fig:2Dbasin} for the CTN dynamics of the ODE system. For the PDE system, we track the dynamical evolution by using pseudospectral methods to discretise in Fourier space and by employing an exponential time differencing method to step in time; see \cite{SAKR16} for details. Starting from different initial guesses for $(z,w)$, taken from a grid in the range $(-0.1,0.1)$ for each $z$ and $w$, we identify the asymptotic states that the PDE system reaches. Figure \ref{fig:2Dpdebasin} shows the types of asymptotic states that are reached when starting from different initial conditions for $(z,w)$. The zero/flat state is when all perturbations decay and the system reaches a quiescent state. When starting with an initial guess of $(z,w)=(0,0)$, we see that the evolution stays at the quiescent state. Regular patterns such as rhombs and hexagons are observed at each of the length scales (wavenumbers) involved, along with more complex patterns that involve both length scales and associated wavenumbers. A dodecagonal quasipattern is observed as the asymptotic state in the PDE system when starting with large values of $w\approx0.05$, irrespective of the value of $z$. The relevant asymptotic state has the value $(z,w)=(-0.0129,0.0669)$ and is the PDE equilibrium associated with $r_4$ which is marked by a black square in Figure \ref{fig:2DODEPDENewton}. All initial conditions  in the $(z,w)$ plane that evolve to this state are indicated by the mustard colour region in Figure \ref{fig:2Dpdebasin}. Here we note that states with small amplitudes, i.e., less than $10^{-4}$, at peaks that arise from higher order combinations of the original wave vectors are still encompassed within this mustard colored region. It is interesting that one of our ODE equilibria is also a relevant potential attractor of the PDE dynamics when starting in a quasicrystal symmetry subspace. 

As discussed using the results in Figs.\,\ref{fig:2DODEPDENewton} and \ref{fig:2DODEPDE}, equilibria of the ODE system generically persist as equilibria for the associated PDE system for parameters close to the original critical point and looking for equilibria with a chosen symmetry. However, the stability characteristics of these equilibria in the PDE system are not straightforward to determine a priori. It may often be the case that the unstable ODE equilibria might be associated with unstable PDE equilibria. Indeed, this has happened for all the unstable ODE equilibria (saddles and maxima) that we have previously considered in the 2-dof ODE system. A more complex situation might be when we have a stable ODE equilibrium, i.e., $r_4$ which is a minimum of the reduced energy landscape and yet the associated PDE equilibrium is highly unstable with multiple eigenvalues that have a positive real part. Since PDE evolutions that are initialised from any initial condition are not restricted to stay within a specific symmetry subspace, many other complex patterns become possible even when we start the evolution within the $(z,w)$ plane. We explore such patterns in further detail in the rest of this section. 

For chosen evolutions that start from near one of the ODE equilibria, we plot the initial condition, the final state and a plot that indicates how much a current state during evolution is like the initial pattern at wavenumber $k=1$ and at wavenumber $k=q$, i.e., we measure the distance between the current PDE state and its projection on the reduced space of wavenumber $k$ and on that of wavenumber $q$. The distance from these projections measures how close (or far) the solution of the PDE lies in comparison to the plane of our 2-dof reduction. In the language of these distances and projections, for the stable PDE state discussed above, we start on the 2-dof plane and we stay on it,
resulting in a stable PDE solution. However, it is also possible that states may be unstable (or even stable) at the 2-dof level, allowing small perturbations in the direction transverse to the $(z,w)$ plane to take advantage of
the infinite degrees of freedom of the PDE to diverge from the plane and result in a different solution. In that light, 
such a plot allows us to visualise how the evolution proceeds to redistribute energy between the length scales (and associated wavevectors) and will allow us to explain our previous observation that initial states with large values of $w\approx0.05$, irrespective of the value of $z$ all converge to the PDE equilibria associated with $r_4$ (shown as a black filled circle), while others as we will see below converge to other states.

\begin{figure}[]
\centering
{\vspace{-0.3cm}
\includegraphics[width=6cm]{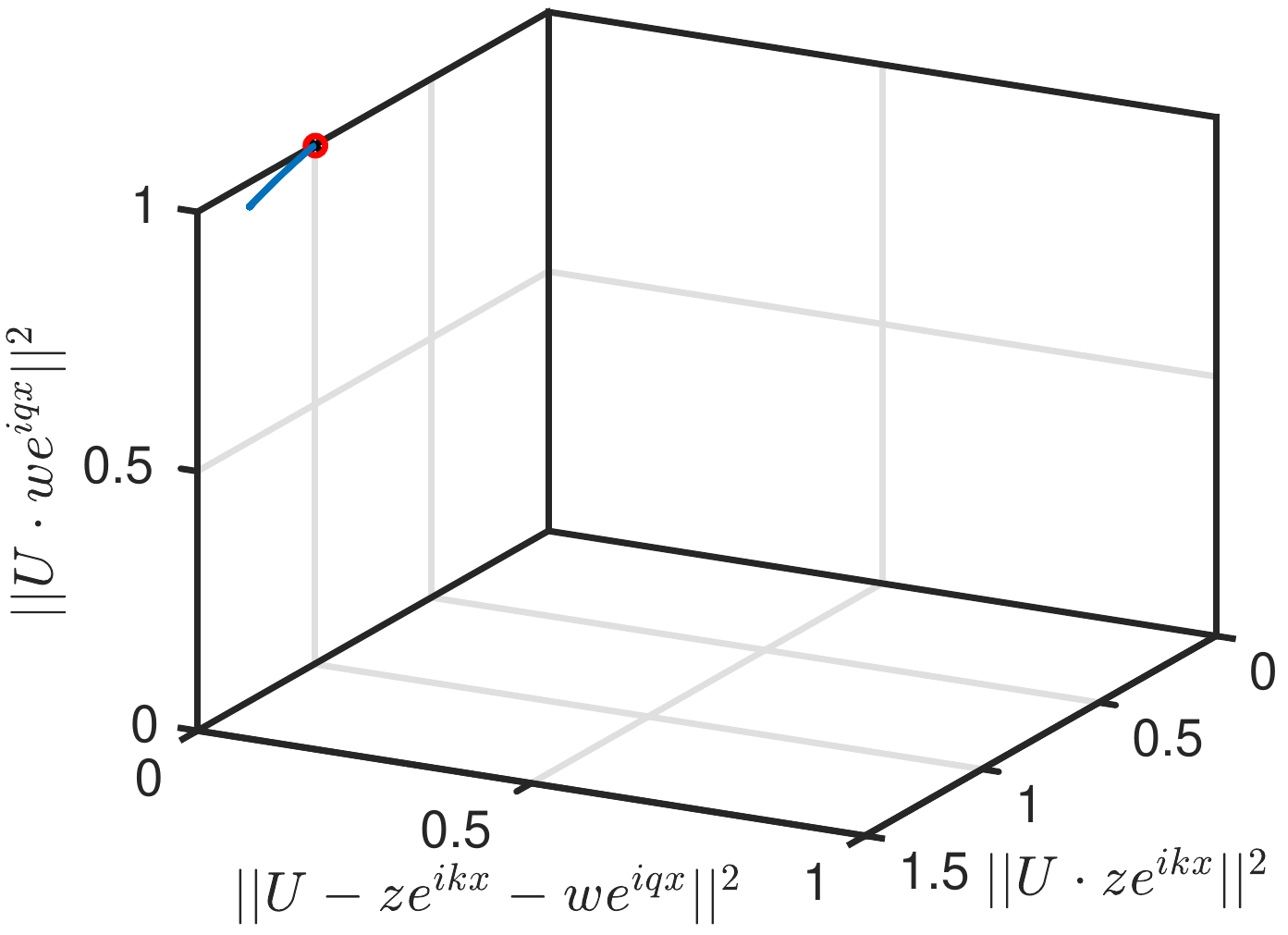}\hspace{0.2cm} \includegraphics[width=3.5cm]{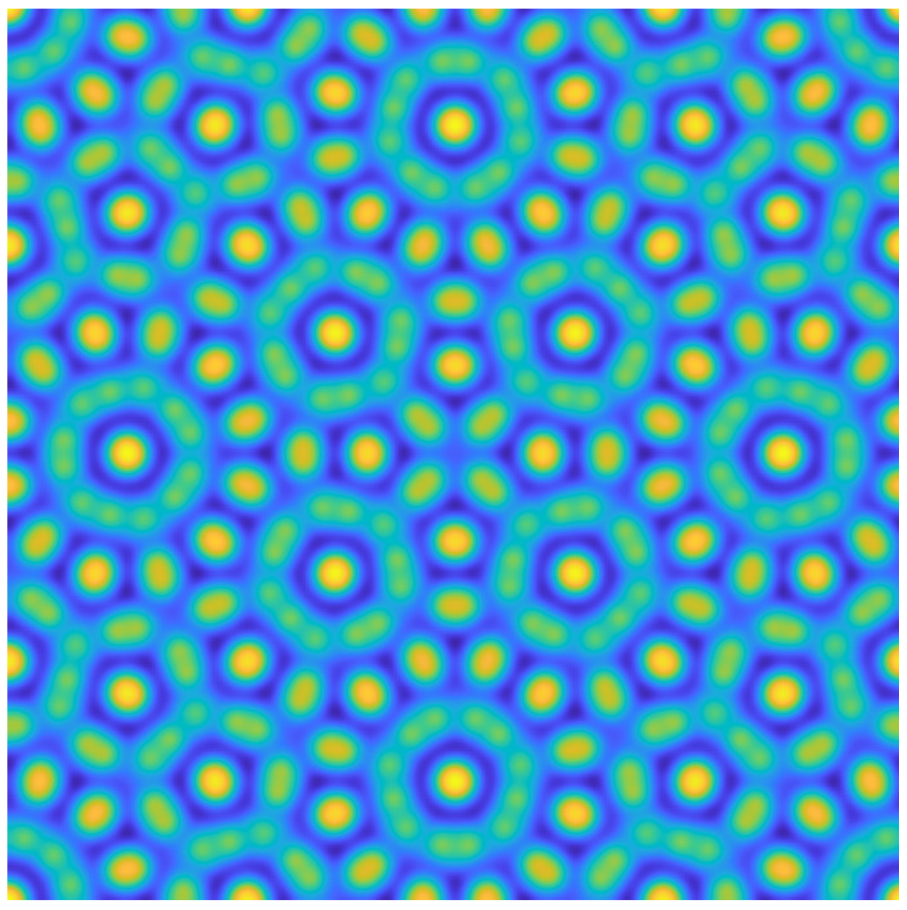}\hspace{0.7cm}\includegraphics[width=3.5cm]{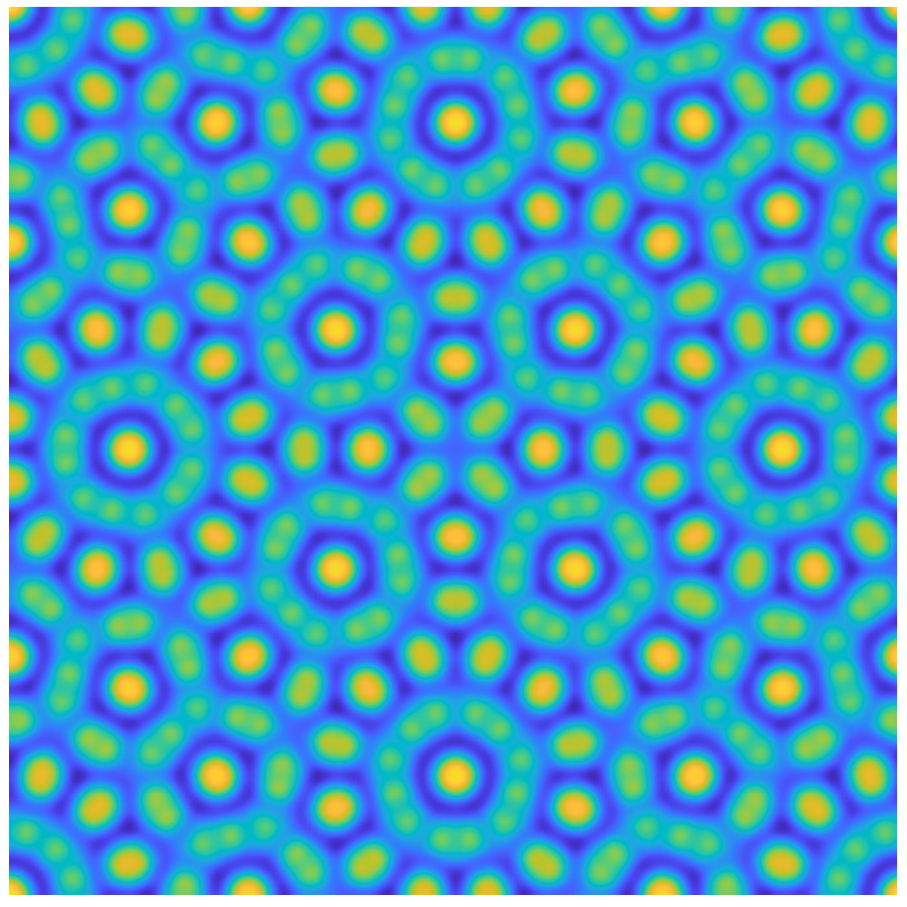}  }
\caption[]{Left panel: Tracking the component of a PDE evolution at the $k=1$ and $k=q$ wavelengths. Time stepping started from amplitudes close to $r4$ ODE solution with $z=-0.01$ and $w=0.07$ (shown as red circle in figure), Middle panel: Prescribed initial condition and Right panel: Asymptotic state.}
\label{fig:r4pdeevol}
\end{figure}

The left panel of Fig.~\ref{fig:r4pdeevol} shows the evolution when we start from a point close to $r_4$, with $(z,w)=(-0.01,0.07)$. We observe that the initial guess, corresponding to the $U$ field shown in the middle panel, has only components along $k=1$ and $k=q$. This can be inferred from the values of the coordinates of the points which are (1,1,0) in the left panel. The value $1$ here refers to a normalised value of the projection of the current state with respect to the wavevectors at $k=1$ and $k=q$ respectively. As time progresses, the evolution moves away from this initial point (identified with a red circle) but stays in the $(z,w)$ plane. The corresponding asymptotic state, shown in the right panel, looks very similar to that of the initial condition. The only change is in slight difference in the value of the amplitudes $(z,w)$ from $(-0.01,0.07)$ to $(-0.0129,0.0669)$, i.e., the convergence to $r_4$. The reduction in both the $(z,w)$ amplitudes from the initial value is reflected in the trajectory of the evolution in the $||U\cdot z e^{ikx}||^2$ and $||U\cdot w e^{iqx}||^2$ plane. 

\begin{figure}[]
\centering
{
\includegraphics[width=6cm]{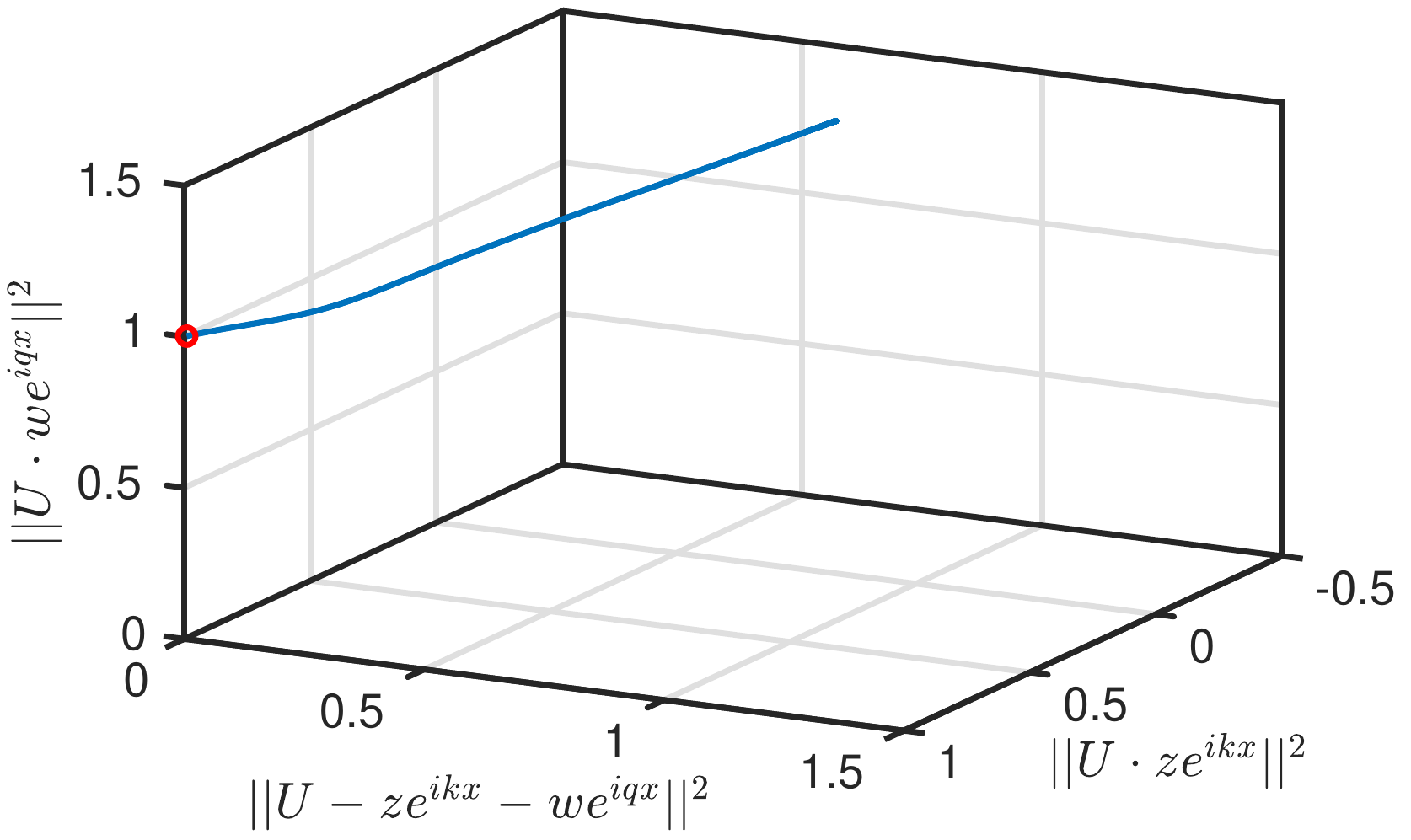}\hspace{0.2cm} \includegraphics[width=3.5cm]{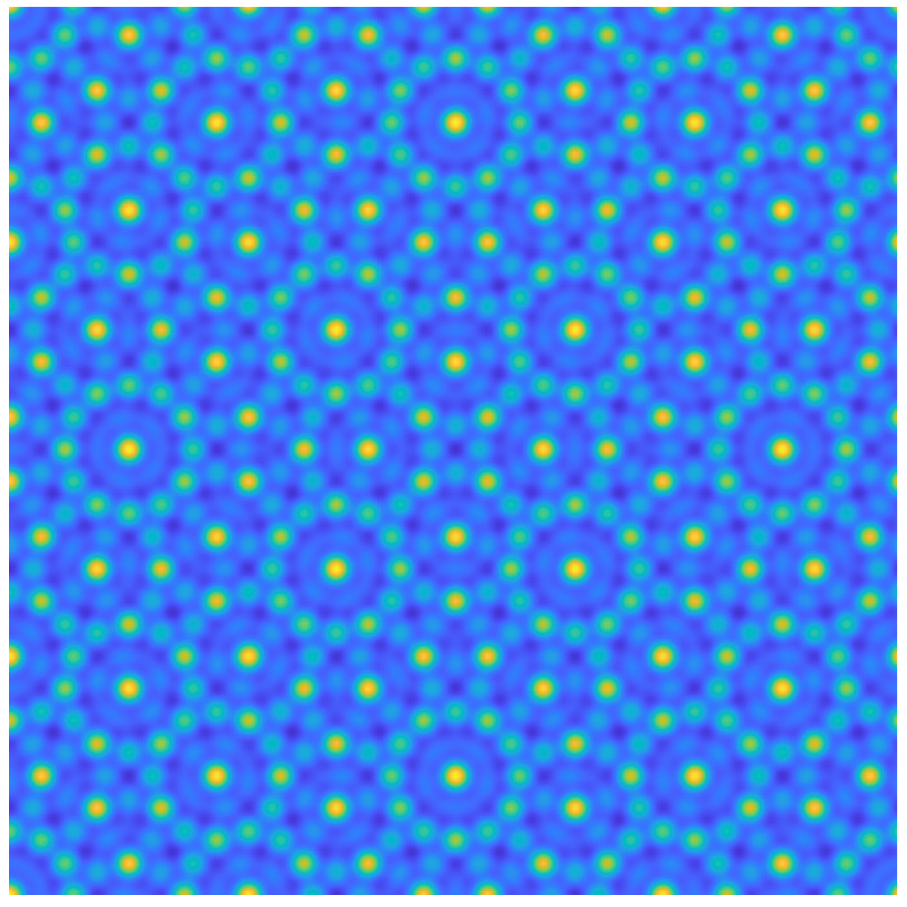}\hspace{0.7cm}\includegraphics[width=3.5cm]{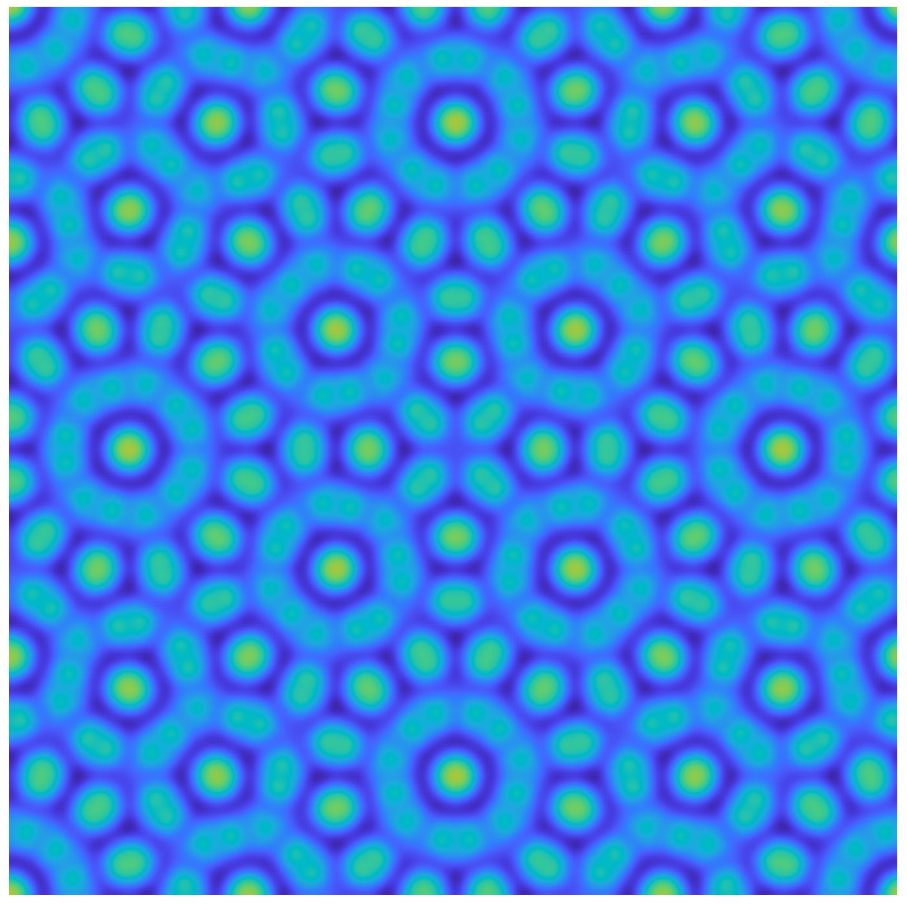}
}
\caption[]{Left panel: Tracking the component of a PDE evolution at the $k=1$ and $k=q$ wavelengths. Time stepping started from amplitudes close to $r6$ ODE solution with $z=0.04$ and $w=0.05$ (shown as red circle in figure), Middle panel: Prescribed initial condition and Right panel: Asymptotic state.}
\label{fig:evcompr6}
\end{figure}

\begin{figure}[]
\centering
{
\includegraphics[width=6cm]{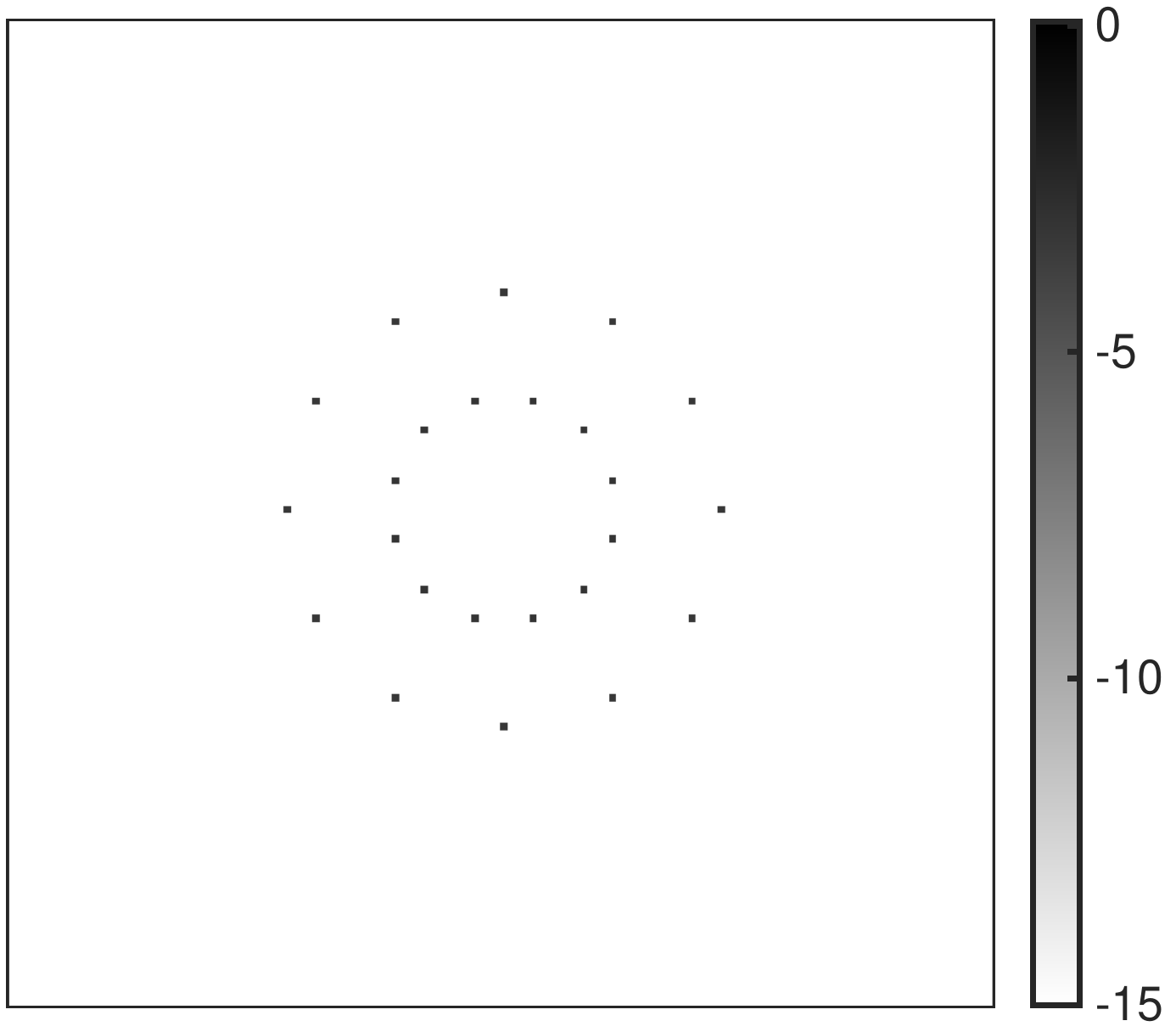}\hspace{1cm}\includegraphics[width=6cm]{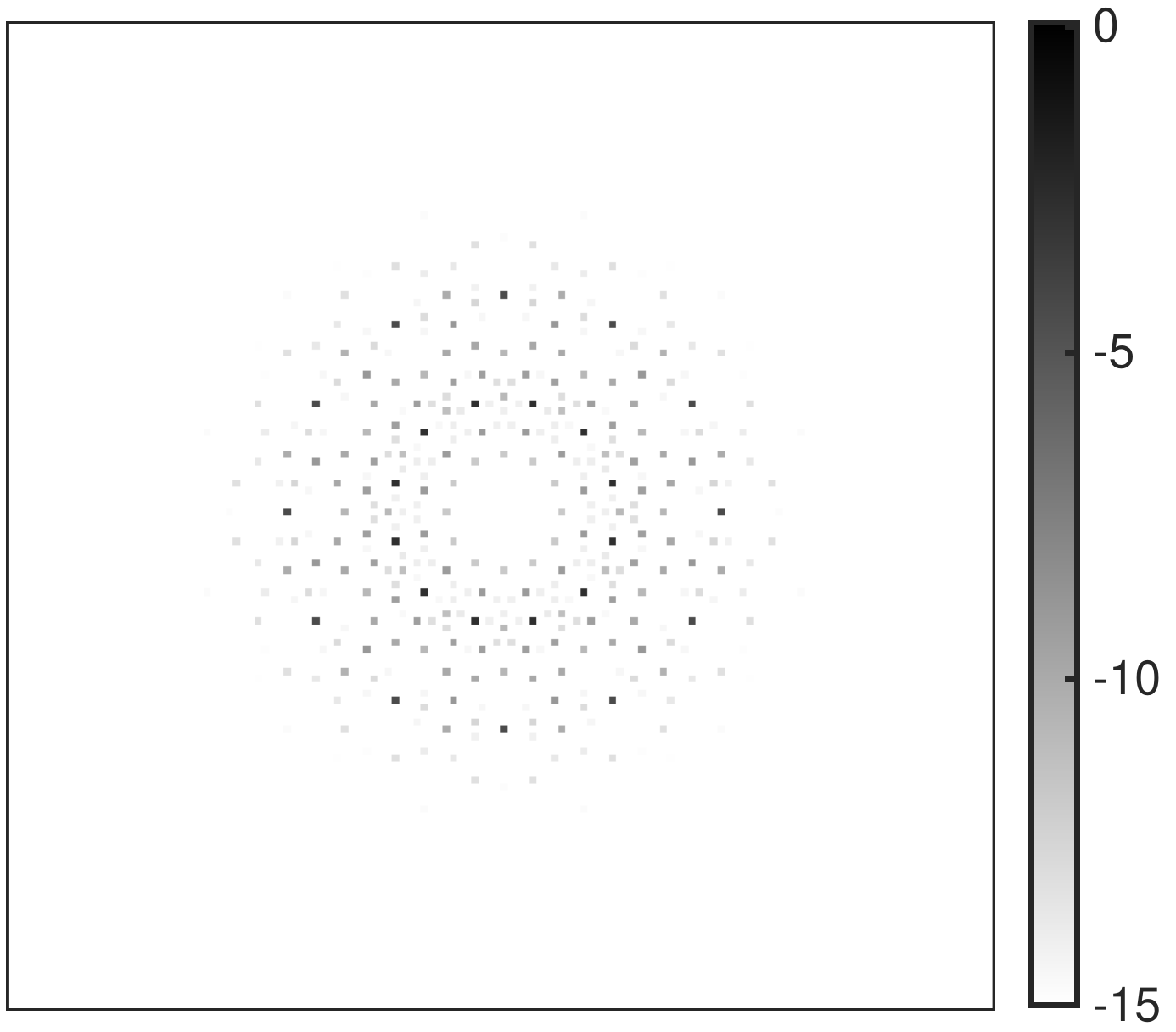}
}
\caption[]{Left panel: Fourier spectra of the prescribed initial condition
  in Fig.~\ref{fig:evcompr6} in log scale and Right panel: Fourier spectra of the asymptotic state also in log scale.}
\label{fig:fftcompr6}
\end{figure}

When starting from an initial guess $(z,w)=(0.04,0.05)$ in Figure~\ref{fig:evcompr6}, which is close to the equilibrium $r_6$ (a saddle for the ODE system), we observe the PDE system to evolve to a different looking quasicrystal pattern. This initial point is within the mustard colored basin of attraction of the PDE equilibrium at $r_4$ (as the classification ignores peaks with amplitudes less than $10^{-4}$ in the asymptotic state). As the phase relation between $(z,w)$ at $r_4$ is not the same as that of the chosen initial condition, we observe that the evolution causes the $z$ amplitude to decrease to negative values, while the $w$ amplitude value increases slightly from the initial guess value.
Furthermore, we look at the Fourier spectra of the initial and final states in Fig. \ref{fig:fftcompr6} plotted in log scale to highlight the differences at small values. Here we see that the initial state has all its amplitudes in one of the modes at either of the two lengthscales. The evolution of the phase between $(z,w)$ adds non-zero peaks at other lengthscales, resulting in the Fourier spectra of the final state having components other than at $k=1$ and $k=q$. This also explains the result in the first panel of Fig. \ref{fig:evcompr6}, where the path tracked ends at an asymptotic state that is out of the $||U\cdot z e^{ikx}||^2$ and $||U\cdot w e^{iqx}||^2$ plane.

\begin{figure}[h]
\centering
{
\includegraphics[width=6cm]{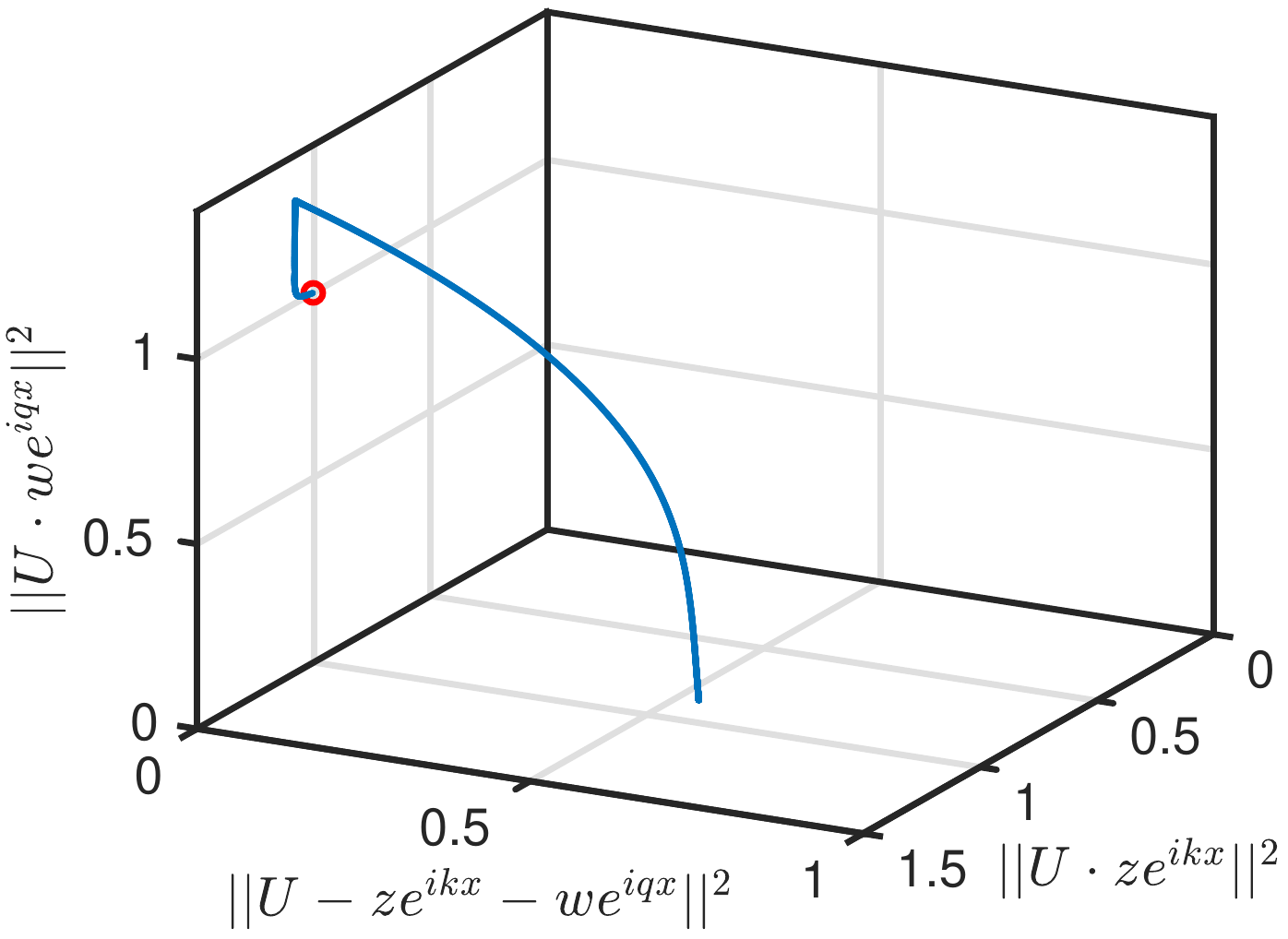}\hspace{0.2cm} \includegraphics[width=3.5cm]{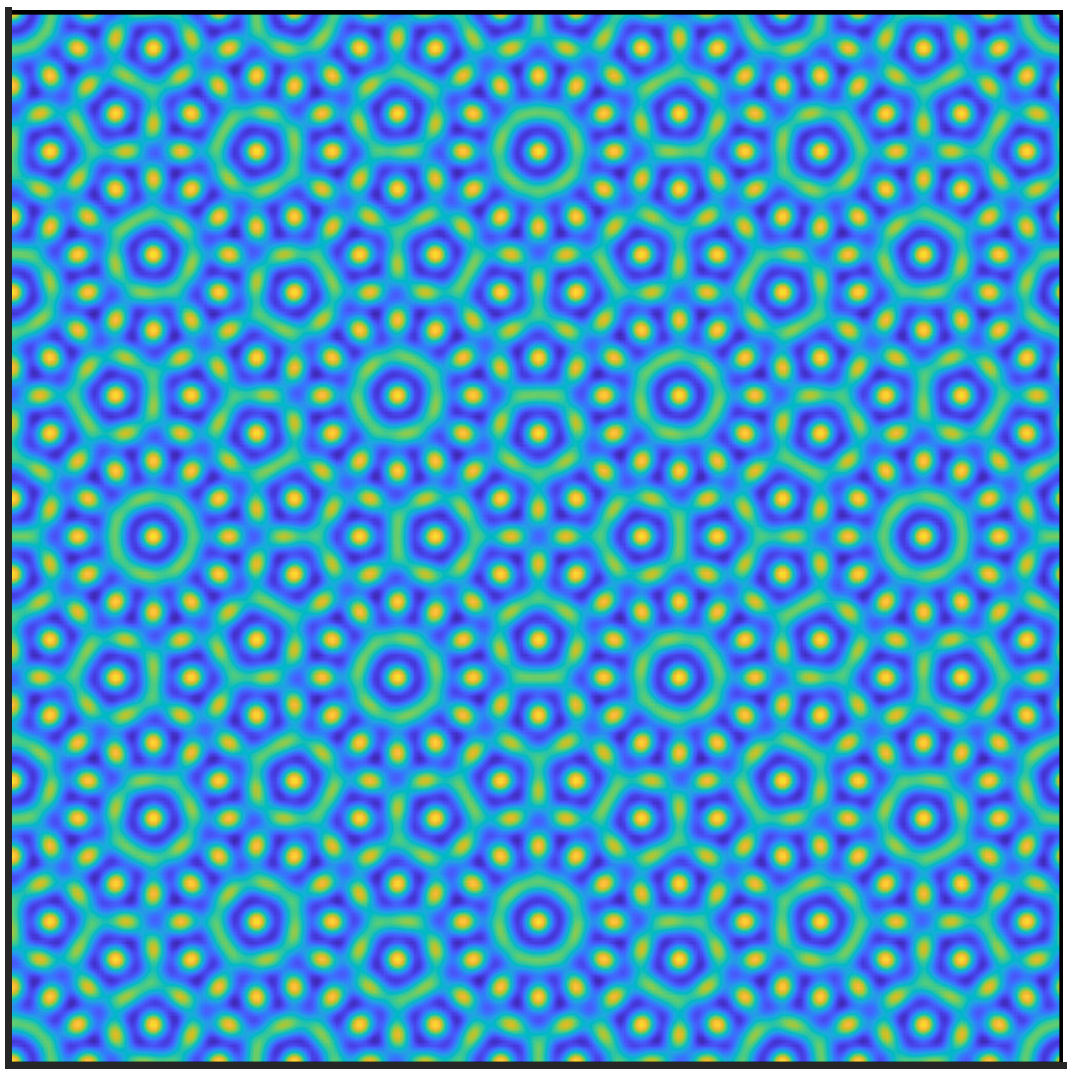}\hspace{0.7cm}\includegraphics[width=3.5cm]{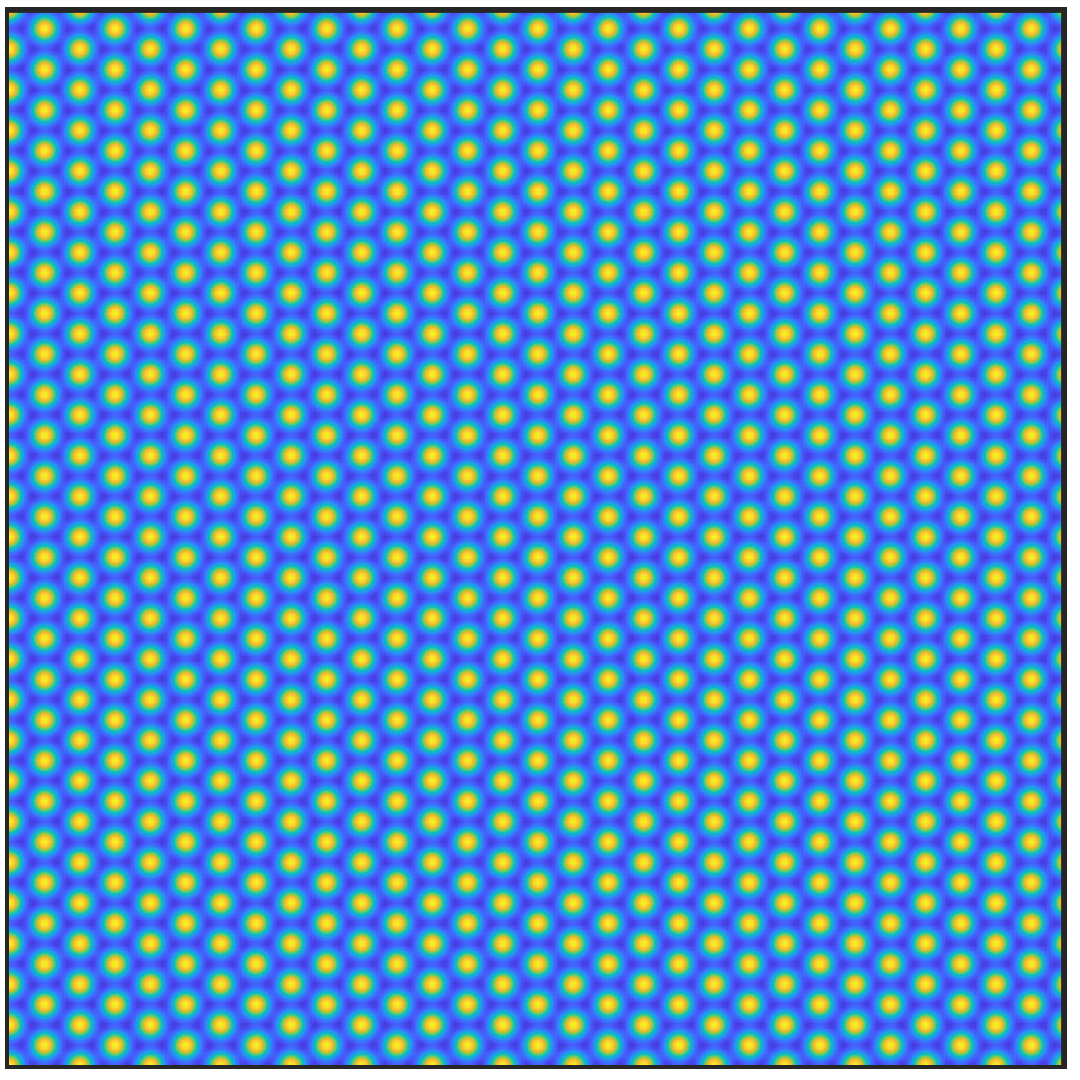}
}
\caption[]{Left panel: Tracking the component of a PDE evolution at the $k=1$ and $k=q$ wavelengths. Time stepping started from amplitudes close to $r9$ ODE solution with $z=0.06$ and $w=-0.01$ (shown as red circle in figure), Middle panel: Prescribed initial condition and Right panel: Asymptotic state. }
\label{fig:evcompr9}
\end{figure}

We then explore in Figure~\ref{fig:evcompr9} an evolution starting from a point close to the $r_9$ equilibrium (which is a minimum for the PDE system) with initial values of $(z,w)=(0.06,-0.01)$. In this case, we see that the evolution evolves along the $(z,w)$ plane for a while as it changes the relative phase between $z$ and $w$, after which there is a sharp change to damp out all the $w$ modes resulting in a $z-$hexagon pattern as the asymptotic state. This is consistent with the initial guess being in the basin of attraction of $z-$hex in Figure \ref{fig:2Dpdebasin}. It is also in line with our discussion about the plane of $(z,w)$ not being an invariant plane constraining the PDE dynamics, but rather the possibility of the PDE to potentially move towards other equilibrium states outside of the relevant plane.

\begin{figure}[h]
\centering
{
\includegraphics[width=6cm]{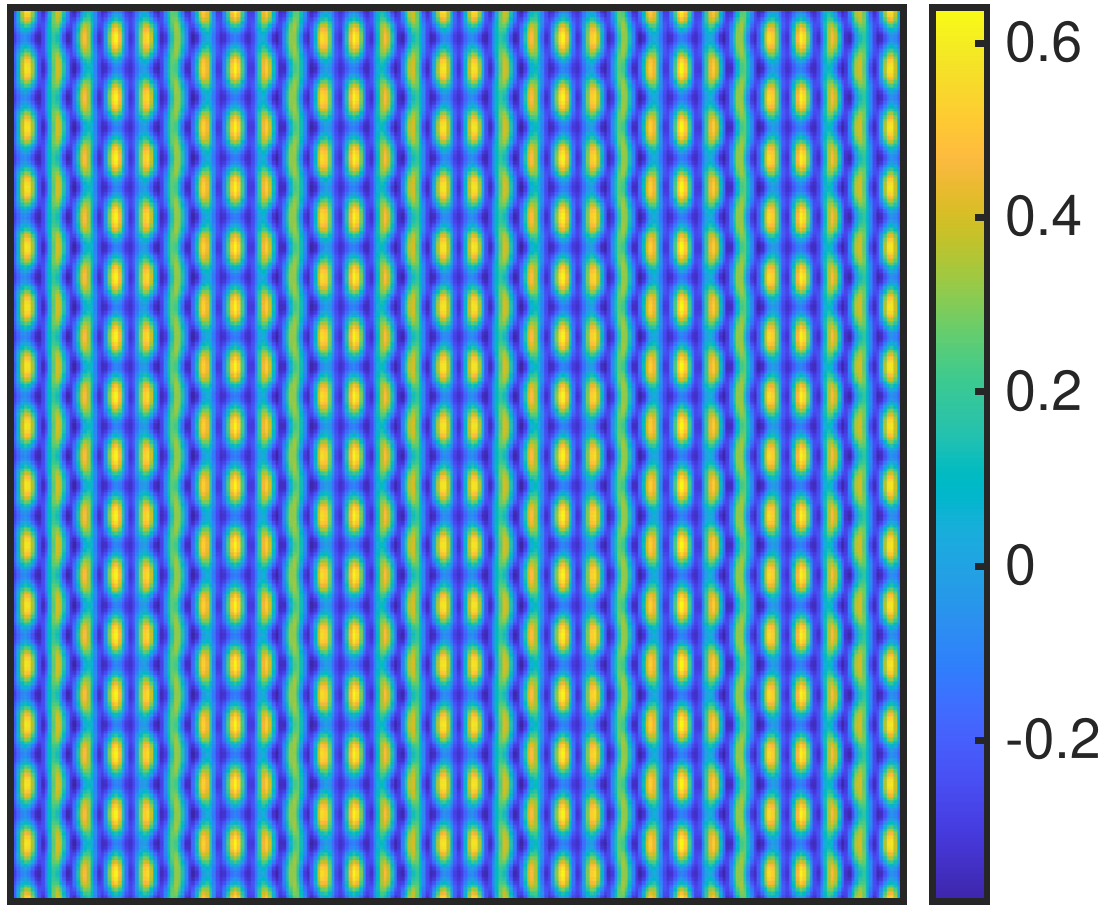}\hspace{0.7cm}\includegraphics[width=6cm]{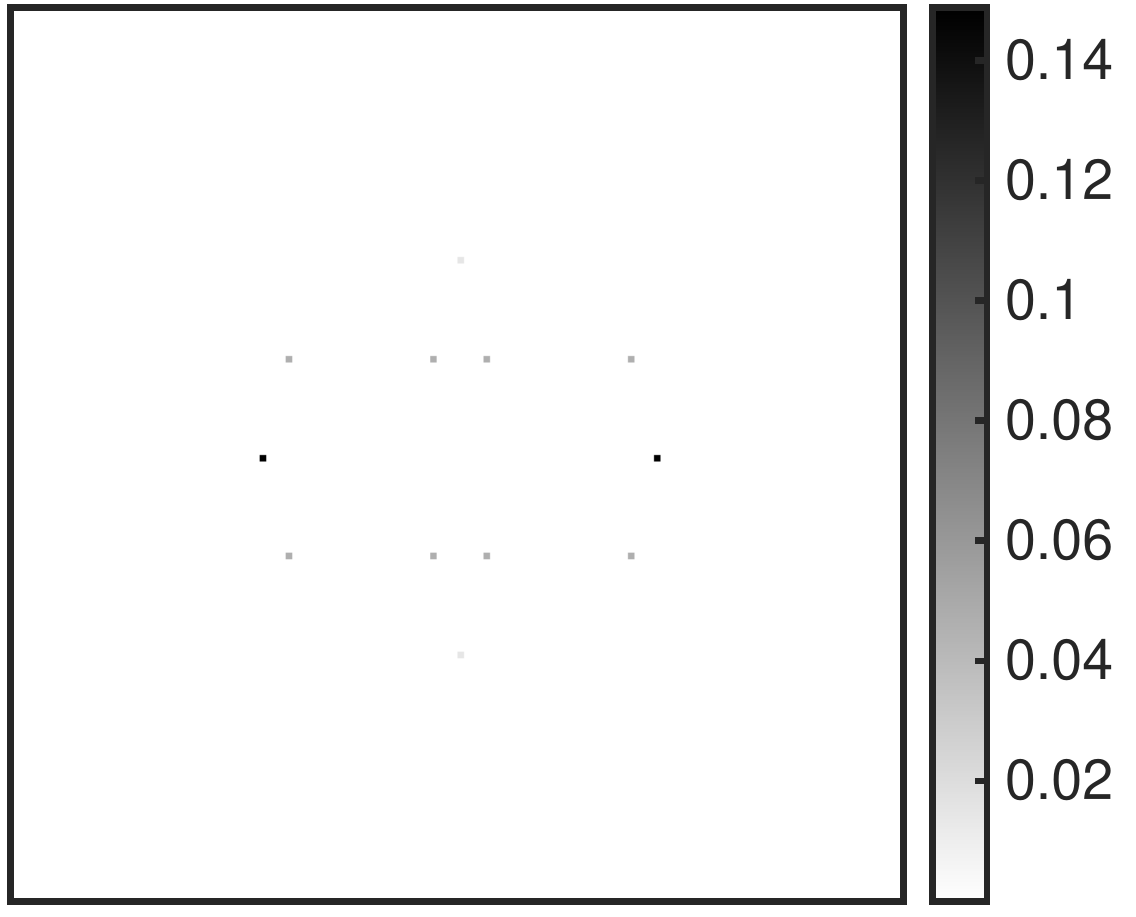}\\
\vspace{0.4cm}
\includegraphics[width=6cm]{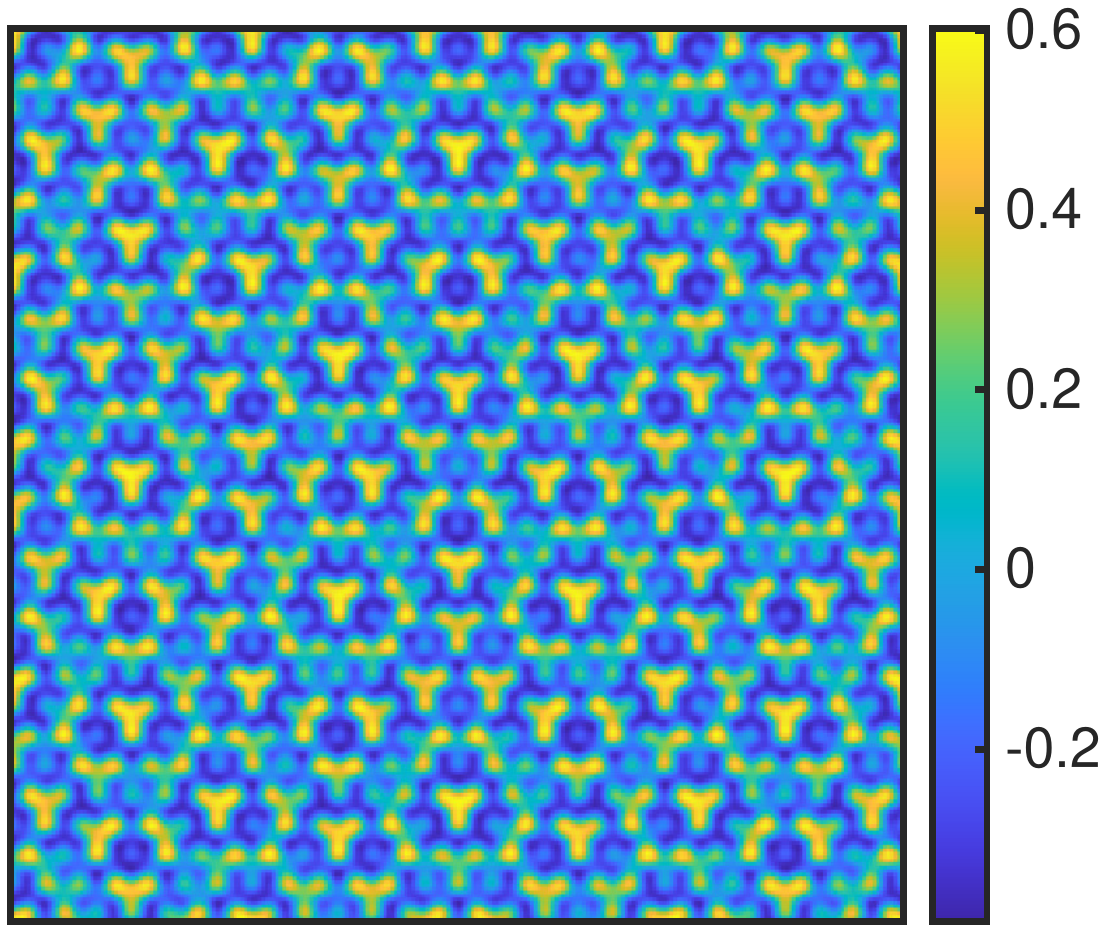}\hspace{0.7cm}\includegraphics[width=6cm]{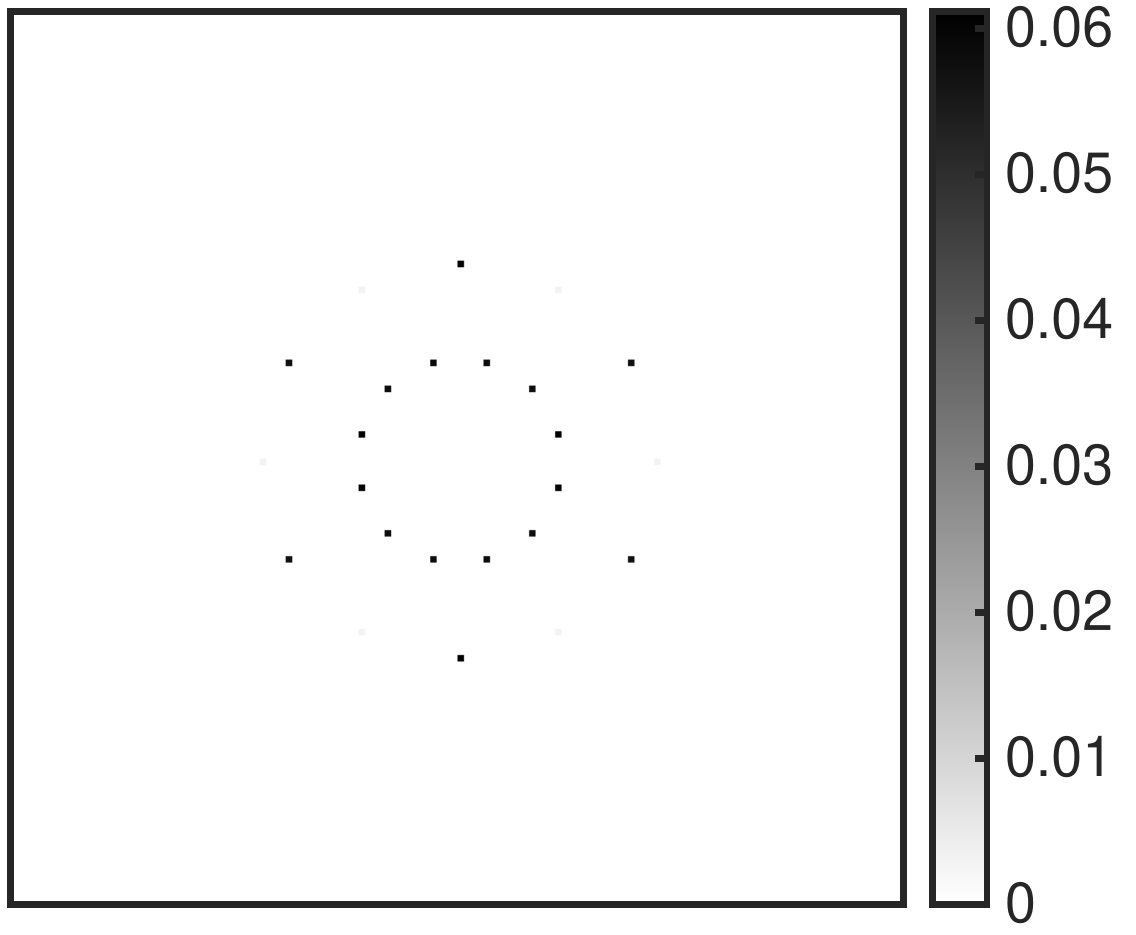}\\
\vspace{0.4cm}
\includegraphics[width=6cm]{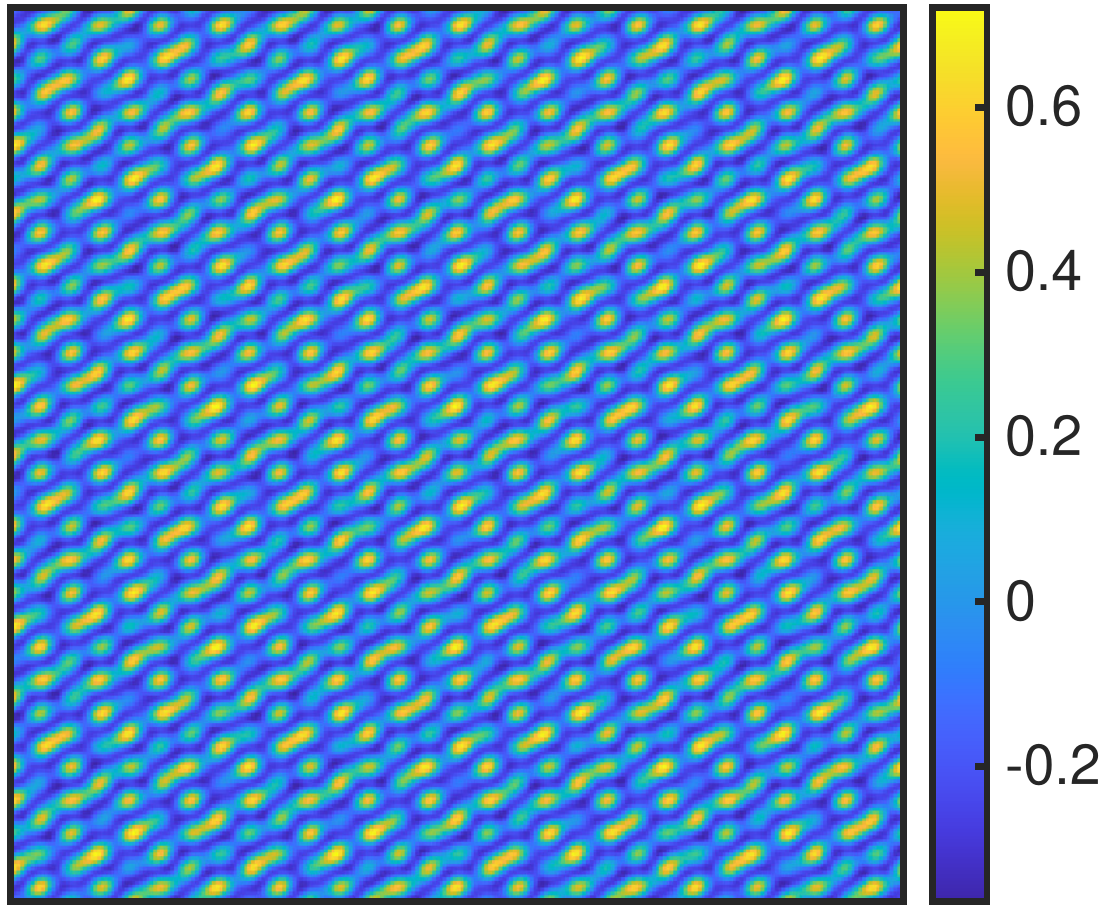}\hspace{0.7cm}\includegraphics[width=6cm]{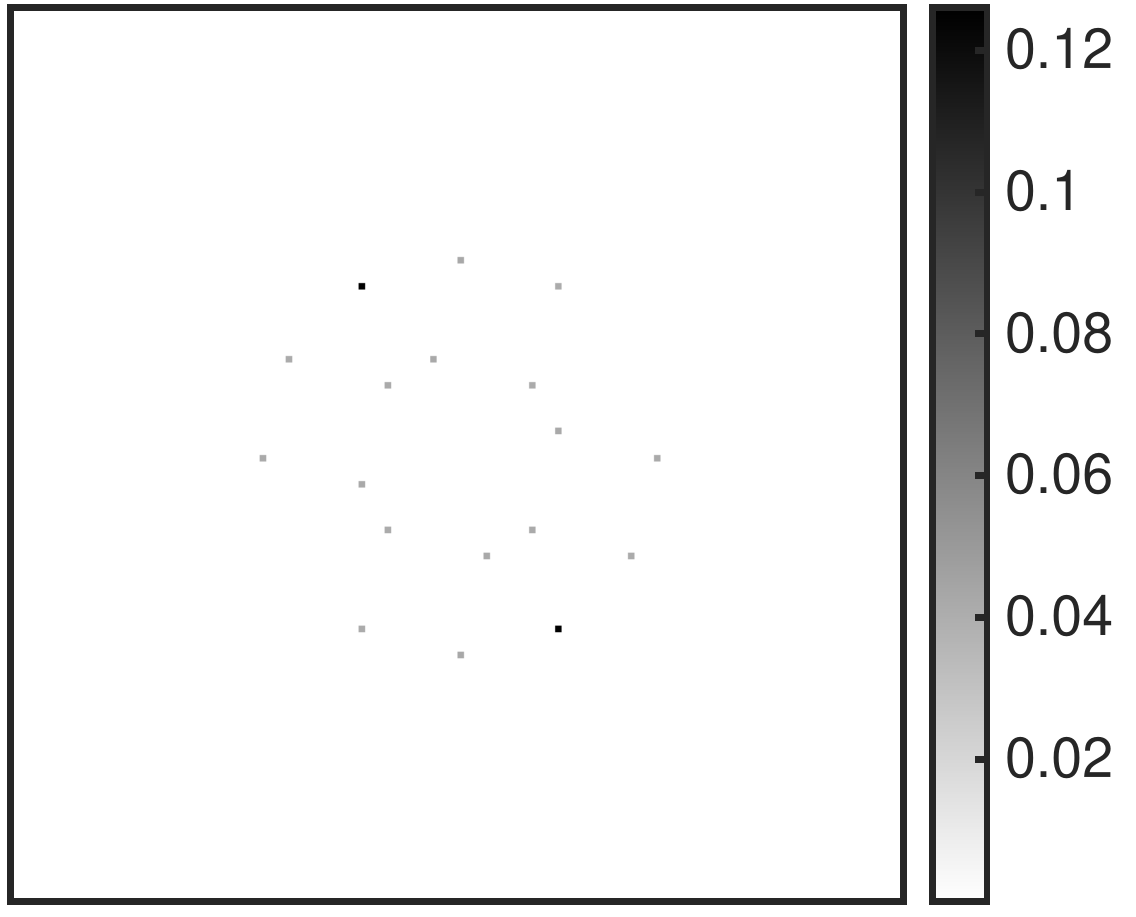}\\
\vspace{0.4cm}
\includegraphics[width=6cm]{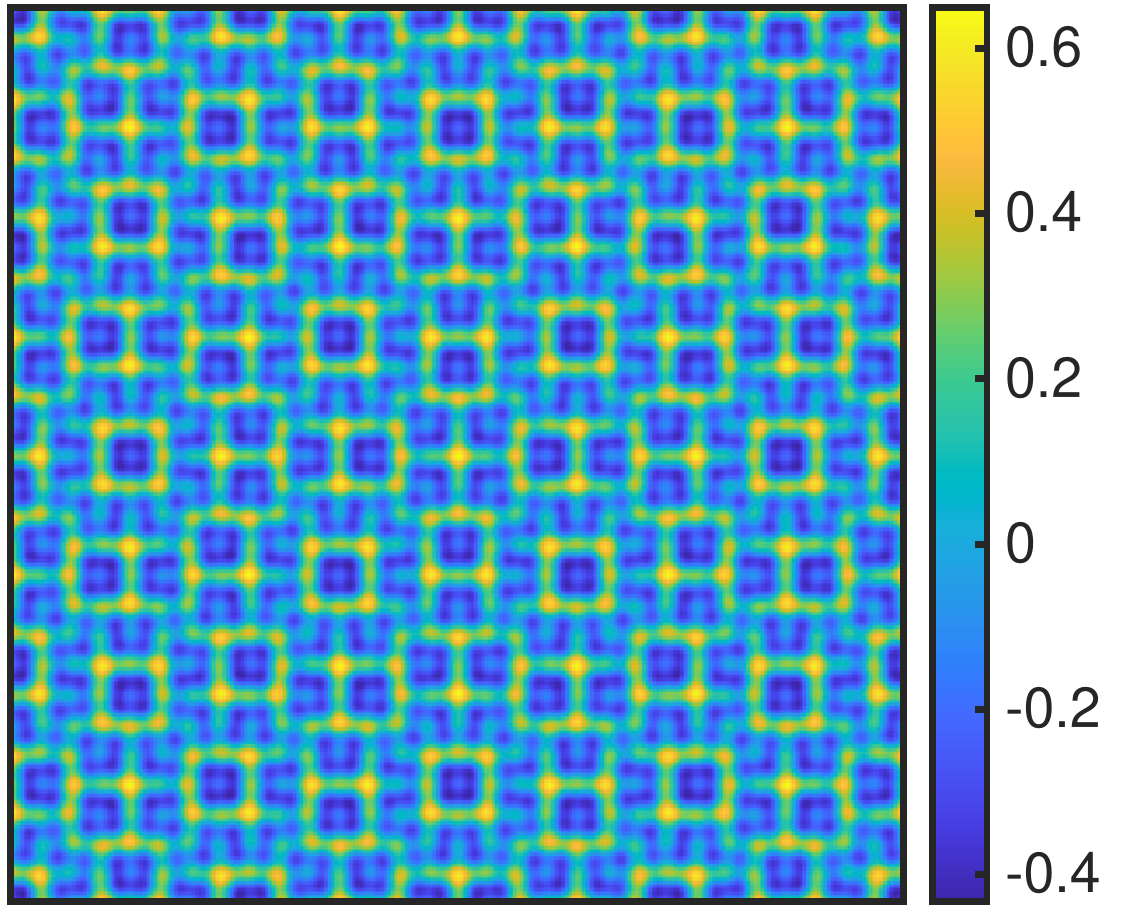}\hspace{0.7cm}\includegraphics[width=6cm]{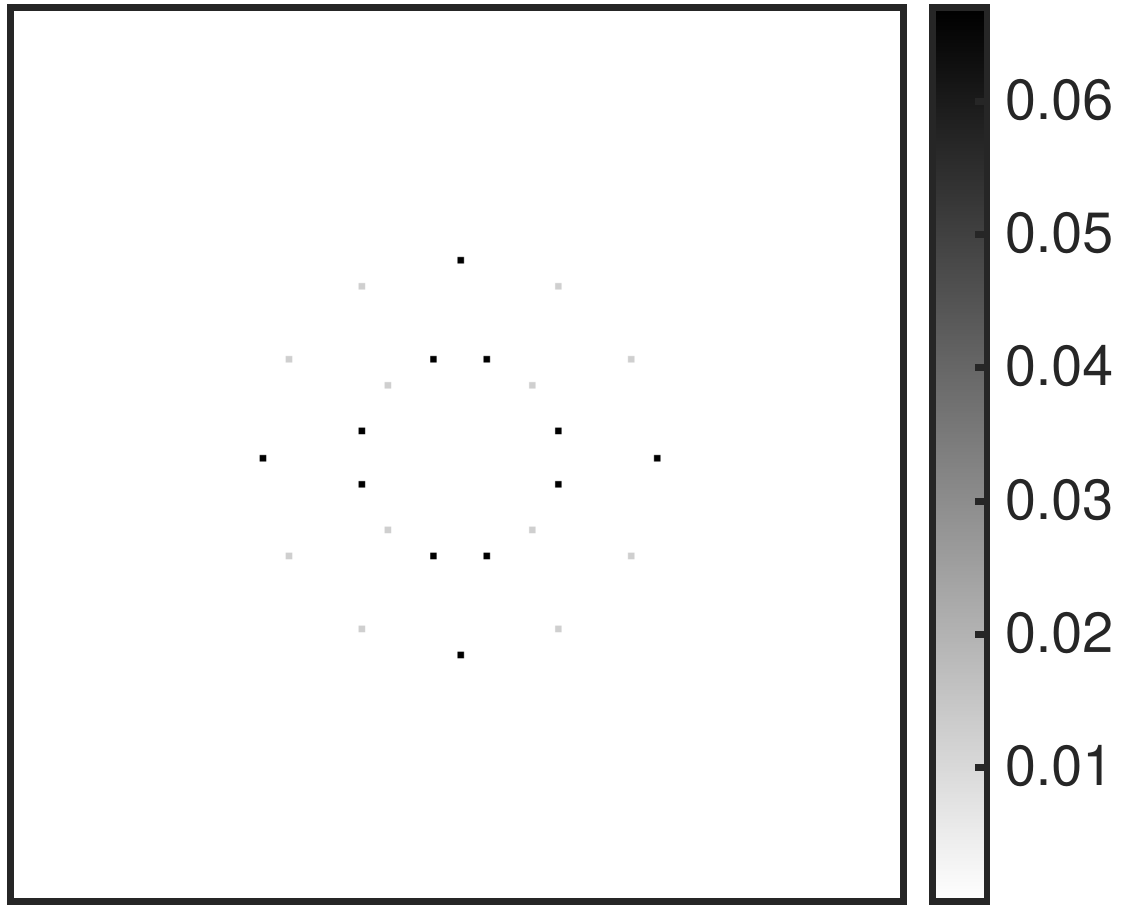}
}
\caption[]{Line 1: Pattern 4i8o and its FFT, Line 2: Pattern 12i6o and its FFT, Line 3: Pattern 8i10o and its FFT, Line 4: Pattern sqsymmQC and its FFT  }
\label{fig:PDEsolns}
\end{figure}

As a final figure, we also include examples of the other types of pattern in the PDE system along with their power spectrum in Figure \ref{fig:PDEsolns}. More details about their symmetry subspaces and the number of unknown amplitudes that need to be determined using ODE reductions to determine them, are listed in Table \ref{tab:eqstates}. The key observation out of this is that the original reduction even via
its unstable solutions enables the exploration of interesting dynamics of
the associated full PDE model. Indeed, our 2-dof plane was the basis for
unraveling a wide variety of PDE-level solutions with different (hexagonal, rhombic, quasi-crystal and other complex pattern) symmetries. Hence, it provides a methodology that combined with flow, Newton and deflation methods allows an in-depth exploration of states accessible to the PDE-problem of interest.

\begin{table}[t]
\caption{Types of equilibria observed as asymptotic states in the PDE system}
\centering
\scriptsize{
\begin{tabular}[b]{@{}C{3cm} c C{3cm}@{} }
\toprule
Type of equilibrium & ($z_1,z_2,z_3,z_4,z_5,z_6; \,\,w_1,w_2,w_3,w_4,w_5,w_6$) & No. of unknown amplitudes  \\
\toprule
Flat & (0,0,0,0,0,0;\,\,0,0,0,0,0,0) & 0 \\
$z$-stripes & (z,0,0,0,0,0;\,\,0,0,0,0,0,0) & 1 \\
$w$-stripes & (0,0,0,0,0,0;\,\,w,0,0,0,0,0) & 1 \\
$z$-hexagons & (z,z,z,0,0,0;\,\,0,0,0,0,0,0) & 1 \\
$w$-hexagons & (0,0,0,0,0,0;\,\,w,w,w,0,0,0) & 1 \\
[2ex]
$z$-rhomb & (0,0,0,z,z,0;\,\,w,0,0,0,0,0) & 2 \\
$w$-rhomb & (0,0,0,z,0,0;\,\,0,0,w,w,0,0) & 2 \\
pattern-12i6o & (z,0,z,0,z,0;\,\,w,w,w,w,w,w) & 2 \\
symm-QC & (z,z,z,z,z,z;w,w,w,w,w,w) & 2 \\
pattern-4i8o & ($z_1$,$z_2$,0,$z_3$,0,$z_2$;\,\,0,0,$w_1$,$w_1$,0,0) & 4 \\
pattern-8i10o & ($z_1$,0,$z_1$,$z_1$,$z_2$,$z_1$;\,\,$w_1$,$w_2$,0,$w_1$,$w_1$,0) & 4 \\
Squaresymm QC & ($z_1$,$z_2$,$z_2$,$z_1$,$z_2$,$z_2$; $w_1$,$w_2$,$w_1$,$w_1$,$w_2$,$w_1$) & 4\\
\bottomrule
\end{tabular}
\label{tab:eqstates}}
\end{table}

\section{Conclusions and Future Challenges}

In the present work, our scope was to develop some simple examples, 
suitable for the testing of different types of methods (the continuous-time
Nesterov method, the square-operator method, the deflation technique,
and of course the widely used Newton's method). As such we started
with a very simple and transparent in its roots, yet sufficiently instructive
1-degree-of-freedom example. There, we saw that flow methods  like CTN can
only lead to local minima. We modified these then to obtain all roots
as local minima using the square-operator method. We appreciated
that neither of these two methods was suitable for utilization in
conjunction with the deflation toolbox. When we turned to Newton's method
plus deflation, we were able to obtain all associated roots.

We then extended consideration to a PDE problem, namely a PFC model for crystallization which is an infinite degree of freedom system. Yet, a reduction can be used at the level of unknown amplitudes at a finite number of wavevectors at two sets of wave numbers that ultimately projects the problem to a 2-degree-of-freedom setting. In such an ODE system, it is possible to obtain all equilibria. It is possible to flow into the stable ones (CTN), flow even to the unstable ones (SOM) and to deflate with Newton to obtain almost all (but not all) starting from the same initial guess. The stability of the different equilibria and the corresponding landscape within the two-dimensional projection was explained, along with the basins of attraction of different methods.

Then, we turned to the full PDE computation. Here, the utilization of Newton iterations enabled the identification of {\it all} the fixed points previously found at the level of the ODE reduction. Working with the same two-dimensional plane of possible initial conditions, but now at the level of the PDE, the dynamical evolution of the model and its potential outcomes were evaluated for arbitrary initial conditions reconstructed from the two-dimensional projection. This allowed us to identify numerous novel PFC patterns and to observe how the dynamics may stay within or depart from the plane of the 2-dof reduction. The eigenvalues at the ODE and PDE levels were compared and similarities/disparities were discussed.

There are numerous important directions to consider for future studies along this vein. In many of the flow problems, a nontrivial issue that arises is that of running into infinities either because of the inherent dynamics of the problem or because of the effective landscape created, e.g., by poles induced by deflation. It would be especially relevant to attempt to generate globally convergent methods that could bypass this type of issue. Indeed, at the level of algebraic numerical tools,
such as Bertini, there exist suitable complex plane tricks~\cite{bertini} that allow for such properties. Comparing methods that are used in that latter context, such as the so-called Davydenko method, with methods here such as the square operator method is also of interest in its own right. It is relevant to note here that if such methods bypassing infinities could be easily built-in, then it may be possible to enable flow methods with deflation to work efficiently
towards identifying all possible roots of a problem. Notice also that deflation here is done ``isotropically'' with  respect to different degrees of freedom. Yet, it is not inconceivable to envision a variant of the method which is anisotropic in its handling of variables. Numerous among these relevant directions are currently under consideration and will be reported in future publications.

\vspace{5mm}

{\it Acknowledgements.} This material is based upon work supported by the US National Science Foundation under Grants No. PHY-1602994 and DMS-1809074 (PGK). PS acknowledges helpful discussions with C\'edric Beaume and Alastair Rucklidge along with support from a Hooke Research fellowship at the Mathematical Institute of the University of Oxford. PGK also acknowledges support from the Leverhulme Trust via a Visiting Fellowship and thanks the Mathematical Institute of the University of  Oxford for its hospitality during part of this work.
 


\end{document}